\documentclass[twocolumn,preprintnumbers,superscriptaddress,nofootinbib,prd,aps,10pt]{revtex4-1}
\pdfoutput=1

\usepackage{graphicx}
\usepackage{latexsym}
\usepackage{amsfonts}
\usepackage{amssymb}
\usepackage{amsmath}
\usepackage{slashed}
\usepackage{array}
\usepackage{dcolumn}
\usepackage{feynmp}
\usepackage{hyperref}

\def\lsim{\mathrel{\raise.3ex\hbox{$<$\kern-.75em\lower1ex\hbox{$\sim$}}}}
\def\gsim{\mathrel{\raise.3ex\hbox{$>$\kern-.75em\lower1ex\hbox{$\sim$}}}}

\usepackage{xcolor}
\definecolor{red}{rgb}{1.0, 0, 0}

\newcommand{ \slashchar }[1]{\setbox0=\hbox{$#1$}   
   \dimen0=\wd0                                     
   \setbox1=\hbox{/} \dimen1=\wd1                   
   \ifdim\dimen0>\dimen1                            
      \rlap{\hbox to \dimen0{\hfil/\hfil}}          
      #1                                            
   \else                                            
      \rlap{\hbox to \dimen1{\hfil$#1$\hfil}}       
      /                                             
   \fi}                                             %

\newcommand{\ra}{\rightarrow}
\newcommand{\gev}{\text{GeV}}
\newcommand{\tev}{\text{TeV}}

\newcommand{\fbinv}{\text{fb}^{-1}}

\newcommand{\charpm}{\tilde{\chi}^\pm}
\newcommand{\charp}{\tilde{\chi}^+}
\newcommand{\charm}{\tilde{\chi}^-}
\newcommand{\neut}{\tilde{\chi}^0}

\begin{document}

\title{Natural Supersymmetry and Implications for Higgs physics}

\author{Graham D. Kribs}
\affiliation{School of Natural Sciences, Institute for Advanced Study, 
             Princeton, NJ 08540}
\affiliation{Department of Physics, University of Oregon,
             Eugene, OR 97403}

\author{Adam Martin}
\affiliation{PH-TH Department, CERN, CH-1211 Geneva 23, Switzerland}
\affiliation{Department of Physics, University of Notre Dame, Notre Dame, IN 46556,
             USA\,}~\thanks{visiting scholar}

\author{Arjun Menon}
\affiliation{Department of Physics, University of Oregon,
             Eugene, OR 97403}

\preprint{CERN-PH-TH-2013-095}

\begin{abstract}

We re-analyze the LHC bounds on light third generation squarks in 
Natural Supersymmetry, where the sparticles have masses  
inversely proportional to their leading-log contributions to the  
electroweak symmetry breaking scale.  Higgsinos are the lightest  
supersymmetric particles; top and bottom squarks are the next-to-lightest
sparticles that decay into both neutral and 
charged Higgsinos with well-defined branching ratios determined
by Yukawa couplings and kinematics.   
The Higgsinos are nearly degenerate in mass,
once the bino and wino masses are taken to their natural (heavy) values.
We consider three scenarios for the stop and sbottom masses:
(I) $\tilde{t}_R$ is light, 
(II) $\tilde{t}_L$ and $\tilde{b}_L$ are light, and 
(III) $\tilde{t}_R$, $\tilde{t}_L$, and $\tilde{b}_L$ are light.
Dedicated stop searches are currently sensitive to Scenarios II and III, 
but not Scenario I\@.  Sbottom-motivated searches ($2 b + \rm{MET}$) 
impact both squark flavors due to $\tilde{t} \ra b \charp_1$
as well as $\tilde{b} \ra b \neut_{1,2}$, 
constraining Scenarios I and III with 
somewhat weaker constraints on Scenario II\@.  The totality of these
searches yield relatively strong constraints on Natural Supersymmetry.
Two regions that remain are:
(1) the ``compressed wedge'', where $(m_{\tilde{q}} - |\mu|)/m_{\tilde{q}} \ll 1$,
and 
(2) the ``kinematic limit'' region, where $m_{\tilde{q}} \gsim 600$-$750$~GeV,
at the kinematic limit of the LHC searches. 
We calculate the correlated predictions for Higgs physics, 
demonstrating that these regions lead to distinct predictions
for the lightest Higgs couplings that are separable with 
$\simeq 10\%$ measurements.  We show that these conclusions remain 
largely unchanged once the MSSM is extended to the NMSSM in order to 
naturally obtain a large enough mass for the lightest Higgs boson 
consistent with LHC data.

\end{abstract}

\maketitle

\section{Introduction}
\label{sec:intro}

Natural Supersymmetry is the holy grail of beyond-the-standard model physics.  
It contains a sparticle spectrum where sparticle masses are inversely 
proportional to their leading-log contributions to the electroweak 
symmetry breaking scale.  
At tree-level the electroweak symmetry breaking scale is determined
by balancing the Higgsino mass-squared against the scalar Higgs mass-squareds.
This implies the leading contribution to electroweak symmetry breaking
comes from the Higgsino mass itself, and thus implies the Higgsinos 
are the lightest sparticles in Natural Supersymmetry.  
The next largest contributions come from one-loop corrections 
from the stops.  We consider the three scenarios: (I) $\tilde{t}_R$ is light, 
(II) $\tilde{t}_L$ and $\tilde{b}_L$ are light, and (III)
$\tilde{t}_L$, $\tilde{t}_R$, and $\tilde{b}_L$ are light.
This spans the space of possibilities for various stop (and sbottom)
mass hierarchies consistent with Natural Supersymmetry.
After this comes the contributions come from the wino and gluino (in the MSSM),
but their masses can be several times larger than the stop masses,
given their comparatively suppressed contributions to electroweak
symmetry breaking.  Natural Supersymmetry suggests the lightest 
electroweakinos can be nearly pure Higgsino-like states.

This spectrum is well-known \cite{Barbieri:1987fn,deCarlos:1993yy,Anderson:1994dz,Cohen:1996vb,Ciafaloni:1996zh,Bhattacharyya:1996dw,Chankowski:1997zh,Barbieri:1998uv,Kane:1998im,Giusti:1998gz,BasteroGil:1999gu,Feng:1999mn,Romanino:1999ut,Feng:1999zg,Chacko:2005ra,Choi:2005hd,Nomura:2005qg,Kitano:2005wc,Nomura:2005rj,Lebedev:2005ge,Kitano:2006gv,Allanach:2006jc,Giudice:2006sn,Perelstein:2007nx,Allanach:2007qk,Cabrera:2008tj,Cassel:2009ps,Barbieri:2009ev,Horton:2009ed,Kobayashi:2009rn,Lodone:2010kt,Asano:2010ut,Strumia:2011dv,Cassel:2011tg,Sakurai:2011pt,Ross:2011xv,Papucci:2011wy,Larsen:2012rq,Baer:2012uy,Espinosa:2012in}, 
and the LHC experiments
have already provided outstanding constraints on simplified models 
involving light stops~\cite{atlas_stops_semi,atlas_stops_had,cms_stops}, 
light sbottoms~\cite{atlas_sbottoms,Chatrchyan:2013lya}, and gluinos that
decay into these
sparticles~\cite{atlas_gluino_new1,atlas_gluino_new2,cms_gluino_new1,cms_gluino_new2}.
Further improvement in the bounds may be possible with specialized 
search strategies, for recent examples see~\cite{Plehn:2010st,Plehn:2011tf,Bi:2011ha,Bai:2012gs,Lee:2012sy,Plehn:2012pr,Alves:2012ft,Han:2012fw,Kaplan:2012gd,Brust:2012uf,Cao:2012rz,Graesser:2012qy,Hedri:2013pvl,Dutta:2013sta,Chakraborty:2013moa,Bai:2013ema}.
However, the results presented thus far typically make strong assumptions
about branching fractions [$BR(\tilde{t}_1 \ra t\,\neut_1) = 1$
or $BR(\tilde{t}_t \ra b\,\charpm_1) = 1$].\footnote{The notable exceptions
are the two recent ATLAS searches for $\tilde{t}_1 \ra t \neut_1$ in
the 1-lepton mode \cite{atlas_stops_semi} 
and all-hadronic mode \cite{atlas_stops_had}, 
where constraints on the branching fraction $BR(\tilde{t}_1 \ra t \neut_1)$ 
were shown, assuming the remaining 
of the branching fraction is unobservable.  Natural Supersymmetry,
however, predicts branching fractions into several channels that
are observable, as we will see.}
In addition, in cases where 
both a light chargino and a light neutralino are present, the results
assume certain mass hierarchies:
$m_{\charpm_1} = 
0.75\,m_{\tilde t_1} + 0.25\, m_{\neut_1}$~\cite{cms_stops} or
$m_{\charpm_1} = 2 \times m_{\neut_1}$~\cite{atlas_stops_semi} or
$m_{\charpm_1} - m_{\neut_1} \gtrsim
50\,\gev$~\cite{atlas_stops_semi}.
These assumptions make it difficult to extract the true bounds on
Natural Supersymmetry.  Consequently, we have undertaken a re-evaluation 
of the constraints on Natural Supersymmetry using the existing LHC results 
on simplified models involving light stops and sbottoms.  

It is also well-known that there is an intricate interplay 
between a light third generation and Higgs physics. 
Supersymmetry predicts the mass of the lightest
Higgs boson to high accuracy through radiative corrections that are 
dominated by just the third generation squarks~\cite{Okada:1990vk,Haber:1990aw,Ellis:1990nz,Barbieri:1990ja,Casas:1994us,Carena:1995bx,Carena:1995wu,Haber:1996fp,Heinemeyer:1998np,Carena:2000dp,Martin:2002wn}. 
If the third generation squarks are
collectively light (say, $\lsim 1$~TeV\@), the predicted mass of
the Higgs boson is too small to be compatible with the 
ATLAS and CMS observation \cite{Aad:2012tfa,Chatrchyan:2012ufa}
of a $125$~GeV Higgs-like boson
(e.g.~\cite{Barbieri:2009ev,Essig:2011qg,Brust:2011tb,Papucci:2011wy,Hall:2011aa,Heinemeyer:2011aa,Arbey:2011ab,Draper:2011aa,Carena:2011aa,Ellwanger:2011aa,Blum:2012ii,Buckley:2012em,Espinosa:2012in,Delgado:2012eu,Carena:2013qia,Carena:2013iba}).
On the other hand, light third generation sparticles can significantly
modify the detailed properties -- production cross section and decay rates --
of the lightest Higgs
boson~\cite{Shifman:1979eb,Spira:1995rr,Kniehl:1995tn,Kileng:1995pm,Kane:1995ek,Dawson:1996xz,Djouadi:1996pb,Djouadi:1998az,Belanger:1999pv,Harlander:2004tp,Dermisek:2007fi,Bonciani:2007ex,Low:2009di,Harlander:2010wr}.

We consider the effects of Natural Supersymmetry on the detailed
properties of the lightest Higgs boson.  Here we are not
interested in \emph{maximizing} a particular decay channel or fitting
to the existing Higgs results, but instead we endeavor to simply understand 
the characteristics 
that Natural Supersymmetry has on Higgs physics.  Our main result is to
overlay the modifications to the Higgs physics onto the allowed parameter 
space of Natural Supersymmetry.  Two interesting regions emerge.
In the ``compressed wedge'' region where $(m_{\tilde{q}} - |\mu|)/m_{\tilde{q}} \ll 1$
and $m_{\tilde{q}}$ can be small, the effects on Higgs physics are to
enhance the inclusive (gluon-fusion dominated) cross section 
$\sigma^{\rm incl}_{\rm MSSM}$ by $10$-$30\%$ simultaneous with a 
slight reduction of $BR(h \ra \gamma\gamma)_{\rm MSSM}$ by up to $5\%$.
By contrast, in the ``kinematic limit'' region where 
$m_{\tilde{q}} \gsim 600$-$750$~GeV, 
there is a slight enhancement of $BR(h \ra \gamma\gamma)_{\rm MSSM}$
by up to $5\%$, with the inclusive (gluon-fusion dominated) cross section 
$\sigma^{\rm incl}_{\rm MSSM}$ within a few $\%$ of the Standard Model
result.  While the experimental situation the LHC collaborations is not 
yet settled, it is already clear that these two regions lead to distinctly 
different effects on Higgs properties that can be probed with $\simeq 10\%$
measurements. 

Given light stops and sbottoms, we must consider the supersymmetric 
prediction for the lightest Higgs boson mass.  
We assert that Natural Supersymmetry -- in the MSSM -- is simply \emph{incompatible} 
with obtaining a lightest Higgs boson mass consistent with the LHC data. 
This point has been emphasized in some recent work, 
for example~\cite{Hall:2011aa,Blanke:2013uia,Barbieri:2013hxa}.
Hence, we do not restrict the third generation 
squark masses to obtain a given lightest Higgs boson mass.  Instead, we assume 
there is another contribution to the quartic coupling that is sufficient
to augment the MSSM contributions, resulting in a Higgs mass that matches
experiment, $m_h \simeq 125$~GeV\@.  Not specifying this contribution
would seem to be fatal flaw of our analysis.  We show that simple extensions
of the MSSM, in particular the next-to-minimal supersymmetric standard model (NMSSM), 
can give both a sufficient boost to the quartic coupling with negligible effects 
on the Higgsino mass spectrum and the decay chains that we consider here.  
Specific examples of NMSSM parameter choices that realize our assertion
are given in Appendix~\ref{sec:NMSSM}.

We do not consider the gluino in this paper.  The gluino contributions to 
the electroweak symmetry breaking scale may be significant in the MSSM, 
given the existing searches that suggest the gluino must be heavier
than $1$-$1.3$~TeV, depending on the search strategy~\cite{Chatrchyan:2013sza,cms_gluino_new1,cms_gluino_new2,atlas_gluino_new1,atlas_gluino_new2}.
However, the size of the gluino contribution to electroweak symmetry
breaking is model-dependent:  A Dirac gluino has a substantially
smaller contribution to the electroweak symmetry breaking scale
compared with a Majorana gluino, when the leading-log enhancements are included, 
allowing a Dirac gluino to be substantially heavier~\cite{Heikinheimo:2011fk,Kribs:2012gx,Benakli:2012cy}.
In addition, the search strategies for a gluino depend on its Majorana or 
Dirac character.  One of the most important search strategies -- 
involving same-sign dileptons (such as~\cite{cms_gluino_new2,atlas_gluino_new1}) 
does not provide a constraint on a Dirac gluino.  \\

\section{Mass Hierarchy in Natural Supersymmetry}
\label{sec:masshierarchy}

\subsection{Contributions to the Electroweak Scale}

In the minimal supersymmetric standard model (MSSM) the electroweak 
symmetry breaking scale is determined by, at tree-level~\cite{Martin:1997ns},
\begin{equation}
\frac{1}{2} M_Z^2 \;=\; \frac{\tan^2\beta + 1}{\tan^2\beta - 1} 
  \frac{m_{H_d}^2 - m_{H_u}^2}{2}
  - \frac{1}{2} m_{H_u}^2 - \frac{1}{2} m_{H_d}^2 - |\mu|^2 \; .
\label{eq:ewsbtree}
\end{equation}
In saying ``contribution to the electroweak scale'', it is understood
that the supersymmetric and supersymmetry breaking parameters
are adjusted to obtain the value already determined by experiment.
Here we are interested in the \emph{relative size} of $|\mu|$ and 
the loop corrections to the electroweak breaking scale, i.e., $M_Z$.

For $\tan\beta$ very near $1$, the coefficient of the first term
in Eq.~(\ref{eq:ewsbtree}) becomes large, because the $D$-flat direction
in the scalar potential is not lifted, and thus implies increased sensitivity
to the supersymmetric parameters.  The sensitivity is most easily understood by 
eliminating dependence on $m_{H_d}^2$ using the tree-level 
relation \cite{Martin:1997ns}
\begin{eqnarray}
m_{A}^2 &=& 2 |\mu|^2 + m_{H_u}^2 + m_{H_d}^2
\end{eqnarray}
to obtain
\begin{equation}
\frac{1}{2} M_Z^2 \;=\; 
  \frac{1}{\tan^2\beta - 1} m_{A}^2
  - \frac{\tan^2\beta + 1}{\tan^2\beta - 1} \left( m_{H_u}^2 + |\mu|^2 \right) \, .
\label{eq:ewsbma}
\end{equation}
At large $\tan\beta$, however, Eq.~(\ref{eq:ewsbma}) simplifies to 
\begin{eqnarray}
\frac{1}{2} M_Z^2 &=& - m_{H_u}^2 - |\mu|^2 \; 
\label{eq:ewsblargetb}
\end{eqnarray}
and eliminates dependence on $m_{A}^2$.  Generally, we have taken
$\tan\beta = 10$ for the analyses to follow, and thus the heavy Higgs
scalars that acquire masses near $m_A$ can be readily decoupled from 
our analysis.  However, the smaller $\tan\beta$ region reappears in our 
discussion of the NMSSM in Appendix~\ref{sec:NMSSM}, where the the 
relative contributions to the electroweak symmetry breaking scale become 
more complicated for the NMSSM scalar potential.

With Eq.~(\ref{eq:ewsblargetb}) in mind, we can compare the relative 
importance of different contributions to the electroweak symmetry
breaking scale by normalizing to $M_Z^2/2$ \cite{Dimopoulos:1995mi} 
\begin{eqnarray}
\Delta( a^2 ) \equiv \left| \frac{a^2}{M_Z^2/2} \right| \; ,
\end{eqnarray}
The tree-level and one-loop contributions are well-known
(e.g., \cite{Martin:1997ns,Papucci:2011wy})
\begin{widetext}
\begin{eqnarray}
\Delta( |\mu|^2 ) &=& 10 \times \frac{|\mu|^2}{(200 \; {\rm GeV})^2} \\
\Delta( \delta m_{H_u}^2|_{\rm stop} ) 
  &=& \frac{3 y_t^2}{8 \pi^2} \left( m_{Q_3}^2 + m_{u_3}^2 + |A_t|^2 \right) 
      \log \frac{\Lambda_{\rm mess}}{(m_{\tilde{t}_1} m_{\tilde{t}_2})^{1/2}}
      \nonumber \\
  &\simeq& 10 \times 
      \frac{m_{Q_3}^2 + m_{u_3}^2 + |A_t|^2}{2 \times (450 \; {\rm GeV})^2}
      \frac{\log \Lambda_{\rm mess}/(m_{\tilde{t}_1} m_{\tilde{t}_2})^{1/2}}{3} \; .
      \label{eq:stopsewsb}
\end{eqnarray}
In the MSSM, there are also important one-loop contributions 
from a Majorana wino and two-loop contributions from a Majorana gluino
(e.g., \cite{Martin:1997ns,Papucci:2011wy})
\begin{eqnarray}
\Delta( \delta m_{H_u}^2|_{\rm wino} )   
  &=& \frac{3 g_2^2}{8 \pi^2} |M_2|^2 \log \frac{\Lambda_{\rm mess}}{|M_2|} 
      \nonumber \\
  &\simeq& 10 \times \frac{|M_2|^2}{(930 \; {\rm GeV})^2}
      \frac{\log \Lambda_{\rm mess}/|M_2|}{3} \\
\Delta( \delta m_{H_u}^2|_{\rm gluino} )   
  &=& \frac{2 \alpha_s y_t^2}{\pi^3} |M_3|^2
  \log \frac{\Lambda_{\rm mess}}{(m_{\tilde{t}_1} m_{\tilde{t}_2})^{1/2}} 
  \log \frac{\Lambda_{\rm mess}}{|M_3|} \nonumber \\
  &=& 10 \times \frac{|M_3|^2}{(1200 \; {\rm GeV})^2}
  \frac{\log \Lambda_{\rm mess}/(m_{\tilde{t}_1} m_{\tilde{t}_2})^{1/2}}{3} 
  \frac{\log  \Lambda_{\rm mess}/|M_3|}{1.5} 
\end{eqnarray}
\end{widetext}
Here we somewhat arbitrarily chose to normalize all of our 
numerical evaluations to a factor of $10$ times $M_Z^2/2$,
as well as normalizing the size of the leading-logs to 
$\Lambda_{\rm mess}/\tilde{m} = 20$.\footnote{Except
for $\Lambda_{\rm mess}/|M_3| = \sqrt{20}$, since a conservative 
interpretation of LHC bounds is that the gluino already exceeds 
$1.3$~TeV in viable scenarios.}
This small ratio implicitly assumes a low scale for the messenger sector,
and thus the smallest sensitivity of supersymmetry breaking parameters
to the electroweak breaking scale. 
This provides suggestive values for $|\mu|$, the stop masses, 
and in the MSSM, the wino and gluino masses. 
As these parameters significantly differ from these suggestive values, 
their relative importance to determining (or fine-tuning to determine)
the electroweak scale is altered accordingly.    
In particular, we see that $|\mu|=200$~GeV gives a comparable 
contribution to a pair of stops at $m_{\tilde{t}_1} = m_{\tilde{t}_2} = 450$~GeV\@.

The Natural Supersymmetry predictions for the wino and gluino mass
depend on whether they acquire Majorana or Dirac masses.
Already we see that if the wino acquires a Majorana mass, its mass 
is expected to be nearly $1$~TeV\@.  A Dirac wino would have a mass
considerably larger.  Similarly a Majorana gluino is expected to be $1.2$~TeV
(with the normalization of the logs as given above), and again
significantly larger than this if it acquires a Dirac mass.

For the purposes of this paper, we assume the gluino is either
sufficiently heavy so as to not lead to collider constraints 
(in practice, this means a Majorana gluino needs to be above about $1.3$~TeV~\cite{Chatrchyan:2013sza, cms_gluino_new1, cms_gluino_new2, atlas_gluino_new1, atlas_gluino_new2}),
or it acquires a Dirac mass, in which case its Natural Supersymmetry
mass is well out of range of the LHC\@.  We assume the wino and bino
acquires $\simeq 1$~TeV masses, but our results are largely insensitive
to this choice.

\begin{figure*}[t!]
\centering
\includegraphics[width=1.0\textwidth]{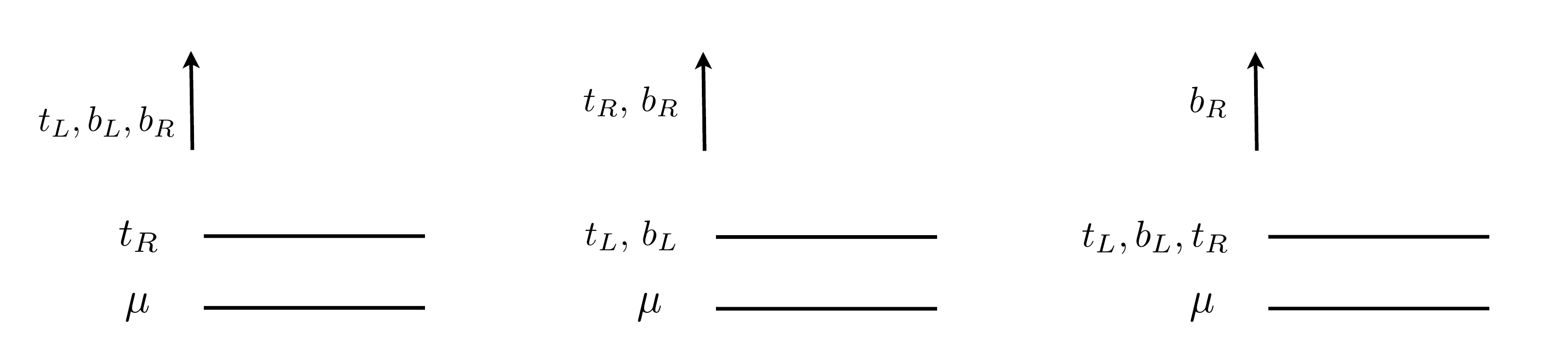}
\caption{The three scenarios of Natural Supersymmetry considered in this paper.
Scenario I,II, and III are illustrated in the left,middle, and right panels.
In our study we varied $\mu \sim m_{\neut_1}$ between $100-500\,\gev$ 
and lightest stop masses $m_{\tilde t_1} \sim 250 - 1000\,\gev$.}
\label{fig:thespectra}
\end{figure*}

\subsection{Higgsino mass splitting}
\label{sec:higgsino_msplit}

In the limit $M_{1,2} \gg |\mu|, v$, the lightest chargino and the lightest
two neutralinos are Higgsino-like and nearly degenerate in mass.
The leading contributions to the mass difference at order $1/M_{1,2}$,
\begin{eqnarray}
m_{\charpm_1} - m_{\neut_1} &=& 
  \frac{M_W^2}{2 M_2} \left( 1 - \sin 2\beta - \frac{2 \mu}{M_2} \right) \nonumber \\
& &{} + \frac{M_W^2}{2 M_1} \tan^2\theta_W (1 + \sin 2\beta) \\
m_{\neut_2} - m_{\neut_1} &=& 
  \frac{M_W^2}{2 M_2} \left( 1 - \sin 2\beta + \frac{2 \mu}{M_2} \right) \nonumber \\
& &{} + \frac{M_W^2}{2 M_1} \tan^2\theta_W (1 - \sin 2\beta) \; ,
\end{eqnarray}
which we can write as
\begin{eqnarray}
m_{\charpm_1} - m_{\neut_1} &=& 
       (3.3 \; {\rm GeV}) \left( \frac{1 \; {\rm TeV}}{M_2} \right) 
       (1 - \sin 2\beta) \nonumber \\
& &{}  + (1.0 \; {\rm GeV}) \left( \frac{1 \; {\rm TeV}}{M_1} \right) 
       (1 + \sin 2\beta) \nonumber \\
& &{}  - (0.6 \; {\rm GeV}) \left( \frac{\mu}{100 \; {\rm GeV}} \right) 
       \left( \frac{1 \; {\rm TeV}}{M_2} \right)^2 \nonumber \\
 m_{\neut_2} - m_{\neut_1} &=& 
       (3.3 \; {\rm GeV}) \left( \frac{1 \; {\rm TeV}}{M_2} \right) 
       (1 - \sin 2\beta) \nonumber \\
& &{}  + (1.0 \; {\rm GeV}) \left( \frac{1 \; {\rm TeV}}{M_1} \right) 
       (1 - \sin 2\beta) \nonumber \\
& &{}  + (0.6 \; {\rm GeV}) \left( \frac{\mu}{100 \; {\rm GeV}} \right)
       \left( \frac{1 \; {\rm TeV}}{M_2} \right)^2 \; . \nonumber
\end{eqnarray}
Clearly, the mass differences among the Higgsinos are just a 
few GeV when $M_{1,2}$ take on natural (heavy) values.
The decays $\charpm_1,\neut_2 \ra \neut_1$ thus yield 
unobservably small energy in the decay products. 
The mass difference is, however, large enough that the decay rates
are prompt on collider time scales and thus there are no macroscopic 
signatures in the detector (at least for wino and bino masses that
do not far exceed $\simeq 1$~TeV).  Hence, for the purposes of LHC detection,
$\charpm_1,\neut_{1,2}$ behave as neutral lightest 
supersymmetric particles that escape the detector as missing energy.

\subsection{Simplified Models of Natural Supersymmetry}

Evidently from Eq.~(\ref{eq:stopsewsb}), the Natural Supersymmetry
prediction for the stop masses depends on the sum
$m_{Q_3}^2 + m_{u_3}^2 + |A_t|^2$.  All other things considered equal,
$A_t \not= 0$ implies the sum $m_{Q_3}^2 + m_{u_3}^2$ must be correspondingly
smaller to hold the sum $m_{Q_3}^2 + m_{u_3}^2 + |A_t|^2$ constant.
We therefore take $A_t$ to vanish.  While this might give some readers pause,
regarding the stop contributions to the lightest Higgs mass, 
recall that we have already asserted that the MSSM is incapable of
providing a sufficient contribution, and so the choice $A_t = 0$
is not inconsistent with our approach.  Instead,
we consider the following mass hierarchies (``Scenarios'') for 
Natural Supersymmetry:
(I) $\tilde{t}_R$ is light, 
(II) $\tilde{t}_L$ and $\tilde{b}_L$ are light, and 
(III) $\tilde{t}_R$, $\tilde{t}_L$, and $\tilde{b}_L$ are light.
These scenarios span the space of possibilities for the stop (and sbottom) 
masses.  We illustrate these scenarios in Fig.~\ref{fig:thespectra}.
The resulting mass eigenstates are given by \cite{Martin:1997ns} 
\begin{eqnarray}
& & \mbox{Scenario I} \;\;\;\;\quad m_{\tilde{t}_1}^2 = 
    m_{\tilde{u}_3}^2 + m_t^2 +   \Delta_{\tilde{u}_R} \\[1em]
& & \mbox{Scenario II} \;\;\quad 
\begin{array}{lcr}
m_{\tilde{t}_1}^2 &=& m_{\tilde{Q}_3}^2 + m_t^2 + \Delta_{\tilde{u}_L} \\[0.5em]
m_{\tilde{b}_1}^2 &=& m_{\tilde{Q}_3}^2 + m_b^2 + \Delta_{\tilde{d}_L}
\end{array} \label{eq:bsquark} \\[1em]
& & \mbox{Scenario III} \;\quad
\begin{array}{lcr}
m_{\tilde{t}_1}^2 &=& m_{\tilde{Q}_3}^2 + m_t^2 + \Delta_{\tilde{u}_L} \\[0.5em]
m_{\tilde{t}_2}^2 &=& m_{\tilde{u}_3}^2 + m_t^2 + \Delta_{\tilde{u}_R} \\[0.5em]
m_{\tilde{b}_1}^2 &=& m_{\tilde{Q}_3}^2 + m_b^2 + \Delta_{\tilde{d}_L} 
\end{array}
\end{eqnarray}
where $\Delta_{\tilde{q}} \equiv (T_q - Q_q \sin^2\theta_W) \cos(2\beta) M_Z^2$.
In Scenario III, we take the soft masses to be equal 
$m_{\tilde{Q}_3} = m_{\tilde{u}_3}$.\footnote{We also include the 
left-right squark mixing contribution $m_t \mu/\tan\beta$, but this is 
suppressed by our choice of $\tan\beta = 10$ as well as $m_t |\mu|$ 
being generally much smaller than $m_{\tilde{Q}_3}^2 = m_{\tilde{u}_3}^2$.}
Since $\cos(2\beta) < 0$ for 
$\tan\beta > 1$, this implies $\Delta_{\tilde{u}_R} > 0$ whereas 
$\Delta_{\tilde{u}_L} < 0$, causing $\tilde{t}_1 \simeq \tilde{t}_L$ and 
$\tilde{t}_2 \simeq \tilde{t}_R$.  Given that we specify soft masses, 
$\tilde{b}_1$ is always lighter than $\tilde{t}_1$ in Scenarios II and III\@.
The mass difference is $(50,30,20)$~GeV for $m_{\tilde{Q}_3} = (200,400,600)$~GeV,
corresponding to $(m_{\tilde{t}_1},m_{\tilde{b}_1}) \simeq
[(260,210),(435,405),(620,600)]$~GeV\@.  Finally, we also 
impose $m_{\tilde{Q}_3,\tilde{u}_3} - m_{\neut_1} > 50$~GeV
for reasons related to the details of the search strategies
employed by ATLAS and CMS.  We discuss this in the next Section.

All other gauginos, all sleptons, and the first and second generation  
squarks are taken to be sufficiently heavy that they do not play a role
in the low energy phenomenology, consistent with Natural Supersymmetry.
We emphasize that the difference between Scenarios I,II, and III 
is not that the stops are believed to be far different in mass,
but simply different enough in mass that the low energy phenomenology
is dominated by one of these three scenarios.


\section{Collider Bounds on Natural Supersymmetry}

\subsection{Collider study setup}
\label{sec:setup_collider}

The inputs for the spectra are the soft masses $m_{Q_3}, m_{u_3}, m_{d_3}$, the $\mu$ term, $\tan{\beta}$, the bottom and top Yukawas, and the weak scale $v$. For simplicity, we will assume the $A_t, A_b$ terms are zero. However, by interpolating between our results for the three given Scenarios, it is possible to reconstruct qualitatively what happens when $A_{t,b} \ne 0$. We set $\tan{\beta} = 10$, and any ``decoupled'' particle in a given Scenario is, for purely practical reasons, taken to have mass $5\,\tev$. Our region of interest is $m_{\neut_1} > 100\,\gev$, $m_{\tilde t_1} > 250\,\gev$. The lower bound on the LSP mass comes from the LEP bound on charginos -- we must obey this bound since $m_{\neut_{1,2}} \sim m_{\charpm_1} \sim \mu$. Finally, while the viability of stops with mass $\lsim 250\,\gev$ remains an interesting question, we concentrate on $m_{\tilde{t}_1} > 250\,\gev$, consistent with the Natural Supersymmetry spectrum.

We also impose an additional restriction on the parameter space, 
namely, to not let the mass difference between the squarks and the Higgsino
become too small.  The experimental analyses on stop production and decay through
$\tilde{t} \ra t \neut_1$ restricted $m_{\tilde{t}_1} - m_{\neut_1} > 175$~GeV 
for ATLAS semi-leptonic and all-hadronic searches
\cite{atlas_stops_semi,atlas_stops_had}, and $> 200$~GeV 
for the CMS semi-leptonic search \cite{cms_stops}.  
Stop decays in compressed spectra often lead to to multiple-body 
final states that are difficult to model without better tools. 
Additionally, the collider limits on 
nearly degenerate spectra become sensitive to how the additional 
radiation in the event (ISR) is modeled.  
We chose to simulate the sensitivity of these searches in 
Natural Supersymmetry for somewhat smaller mass differences,
$m_{\tilde{Q}_3,\tilde{u}_3} - m_{\neut_1} > 50$~GeV\@.
As we will see, we find the existing LHC searches are sensitive 
to Natural Supersymmetry with splittings this small.
However this region needs to be interpreted with some care,
since we are obtaining constraints from both stop searches and
sbottoms searches.  Sbottom searches have somewhat different
restrictions on the kinematics, but since we chose a minimum mass 
difference between the soft mass $m_{\tilde{Q}_3}$ and the lightest neutralino,
the highly compressed region with respect to the sbottom and neutralino 
is not simulated.  We therefore do not anticipate significant changes 
in the bounds for the parameter space we consider.

For each scenario at a given mass point ($\mu$, $m_{\tilde{q}}$), events are generated using PYTHIA6.4~\cite{Sjostrand:2006za}. We use CTEQ6L1~\cite{Pumplin:2002vw} parton distribution functions and take all underlying event and multiple interaction parameters to their values specified in Ref.~\cite{Chatrchyan:2013sza,mrenna}.  The cross section is calculated by summing the next-to-leading order plus next-to-leading log (NLO + NLL) values \cite{Beenakker:1996ch,Beenakker:1997ut,Kulesza:2008jb,Kulesza:2009kq,Beenakker:2009ha,
Beenakker:2010nq,Beenakker:2011fu}
using~\cite{susyxsec} over all light ($< \tev$) 3rd generation sparticles in the spectrum. Following PYTHIA generation, the events are fed into DELPHES~\cite{Ovyn:2009tx} to incorporate detector geometry and resolution effects. We use the default DELPHES ATLAS and CMS detector descriptions, but modify the jet definitions to agree with the corresponding experiment: anti-$k_T$ algorithm, with size $R = 0.4$ for ATLAS and $R = 0.5$ for CMS analyses. Additionally, while the experimental flavor-tag/fake-rates slightly differ from analysis to analysis, we used a fixed $70\%$ tag rate for all $b$-jets that lie in the tracker ($|\eta_j| < 2.5$).

The signal simulation is used to derive the efficiency --  the survival rate in a given bin of a particular analysis. Given our simplified supersymmetry spectra, this efficiency is a function of the squark and LSP masses alone, i.e. for bin $i$ we find $\epsilon_i(m_{\tilde t_1}, m_{\neut_1})$. The product of the derived efficiency with the luminosity and the  cross section (at NLO + NLL) is the number of signal events, $s_i$.
\begin{equation}
s_i = \mathcal L \times \sigma_{\text{NLO+NLL}}(m_{\tilde t_1}) \times \epsilon_i (m_{\tilde t_1}, m_{\neut_1}). \nonumber
\end{equation}
 We derive exclusion limits by comparing $s_i$ at a given $(\tilde m_1, m_{\neut_1})$ with the number of signal events allowed at 95\% CL calculated with a likelihood-ratio test statistic. Specifically, the 95\% CL limit on the number of signal events, $s_{i,95}$ is the solution to
\begin{equation}
0.05 = \frac{\Pi_i \text{Pois}(n_i|b_i + s_{i, 95})}{\Pi_i \text{Pois}(n_i | b_i) }
\end{equation} 
where the $n_i$ is the number of observed events in a channel $i$, $b_i$ is the number of expected SM background events, and we take the product over all orthogonal channels. We take both $n_i$ and $b_i$ directly from the experimental  papers. 

To incorporate systematic uncertainties, the number of background events in a bin is allowed to fluctuate: $b_i \rightarrow b_i\,(1 + \delta b_i)$. After multiplying by a Gaussian weighting factor, we integrate over $\delta b_i$, following Ref.~\cite{conway}. We take the width of the Gaussian weighting factor to be the relative systematic uncertainty in a given bin quoted by the experiment, $f^b_i$, using the larger error if asymmetric errors are given\footnote{The Gaussian integration is truncated such that the number of background events is always positive.}.
\begin{align}
\text{Pois}(& n_i | b_i + s_{i, 95})  \rightarrow  \nonumber \\
& \int \delta b_i\, \text{Gaus}(\delta b_i, f^b_i)\text{Pois}(n_i | b_i\,(1 + \delta b_i) + s_{i,95})
\end{align}

One may ask if a likelihood-ratio analysis is really needed, instead of just a rescaling of existing bounds. If the signal yield according to ATLAS/CMS was given for  each $(m_{\tilde t_1}, m_{\neut_1})$ bin, then we could rescale and determine the yield, and thereby the exclusion bounds, in each of our Scenarios. However, such detailed information is not public  and only the signal yields at specific benchmark points are given. In order to extrapolate yields away from the benchmarks, some model is needed, and for that we rely on the simulation method described above.

Before describing the details of the searches we consider, it is important to emphasize that the {\em absolute} bounds we present are only approximate. To derive the signal efficiency we have used fast-simulation tools (DELPHES) whose energy smearing and tagging functions are approximations -- usually optimistic -- of the full detector effects. In multi-jet, especially multi-$b$-jet final states, the differences between the fast and full-detector simulations add up, making it tricky for us to match the quoted absolute bounds on a given scenario. To improve the accuracy of the absolute bounds, the Scenarios presented here should be studied by CMS/ATLAS themselves, either as a dedicated reanalysis or using a tool such as RECAST~\cite{Cranmer:2010hk}. Meanwhile, the relative bounds, i.e. the difference between Scenario I and Scenario II, are robust.

\subsection{Direct stop searches}
\label{sec:dstop}
In this section we present the limits on the supernatural scenarios from the most recent LHC direct stop searches~\cite{cms_stops, atlas_stops_semi, atlas_stops_had}. Bounds from these searches are usually (though not always) cast in term of stops that decay either 100\% of the time to a top quark and a neutralino or 100\% of the time to a bottom quark and a chargino.  Stops are searched for in several different final states, and the first two stop analyses we consider are semileptonic searches. While the details differ between the ATLAS and CMS searches (see Appendix \ref{app:search} for the full analyses description), both require a hard lepton, significant missing energy, and at least four jets, one of which must be tagged as a $b$-jet.

\begin{figure}[t!]
\centering
\includegraphics[width=0.48\textwidth]{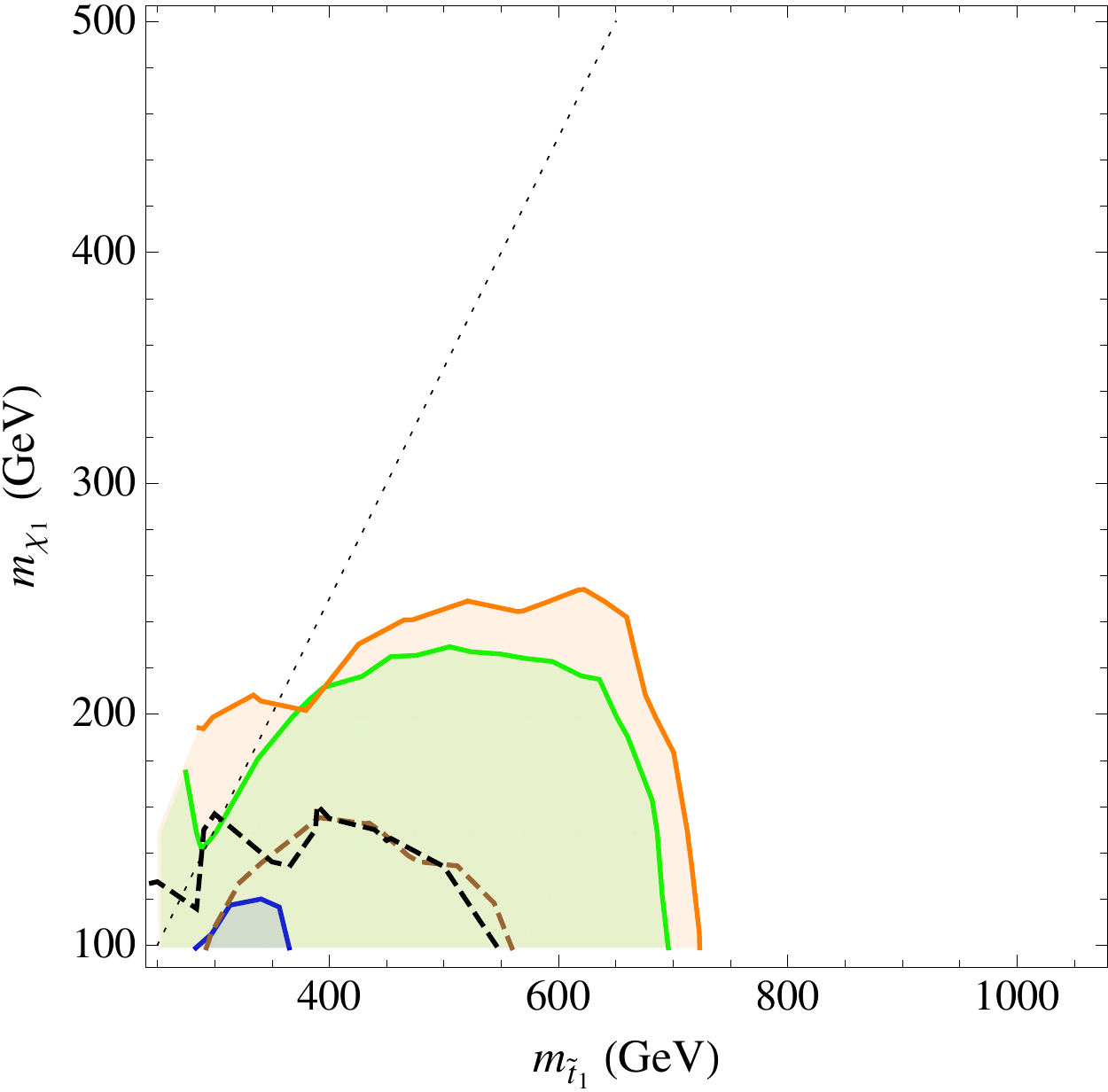}
\caption{Limits on the various 3rd generation scenarios coming from the CMS direct stop search (semileptonic channel). The $x$-axis corresponds to the lightest physical stop mass $\tilde{t}_1$ in each Scenario.  In Scenarios II and III, one $b$-squark is present with a physical mass that is slightly lighter, given by Eq.~(\ref{eq:bsquark}). The orange contour shows the 95\% exclusion bound on Scenario III, the green contour is the bound on Scenario II, and there is no bound on Scenario I\@. The black, dashed contour is the bound from this analysis using CMS simplified model T2tt that involves direct production of stops which decay solely to unpolarized top plus neutralino; $\tilde t_1 \rightarrow t + \neut_1$. The brown dashed line shows the limit on a second default scenario: right-handed stops decaying to top plus bino. The difference between the black and brown dashed lines gives some indication how the polarization of the top can affect limits. The black dotted line is $m_{\tilde{t}} - m_{\neut_1} = 150\,\gev$, which is the self-imposed restriction on the CMS analysis, since ISR is not properly taken into account when the signal is generated with PYTHIA. We have also restricted our re-analysis in the ``compressed wedge'' region [where $(m_{\tilde{q}} - m_{\neut_1})/m_{\tilde{q}} \ll 1$], requiring $m_{\tilde{Q}_3} - m_{\neut_1} > 50$~GeV, that results in the excluded region extending slightly to the left of the black dotted line. See the text for details.}
\label{fig:cmsstops}
\end{figure}

Running our three Scenarios through the CMS direct stop search~\cite{cms_stops}, we find the following exclusion contours in the $(m_{\tilde t_1},\, m_{\neut_1})$ plane (Fig.~\ref{fig:cmsstops}). This is somewhat an abuse of notation -- the horizontal axis actually corresponds to the mass of the lightest stop eigenstate for a given spectra\footnote{For example, in Scenario III the spectrum also contains a second stop and a sbottom -- all three states are produced and analyzed when deriving the analysis efficiencies, though limits are still placed in terms of the lightest stop eigenstate}. For comparison, we include limits from two  ``default'' spectra (calculated in the same manner as our three Scenarios): 
\begin{enumerate}
\item Stop production and decay with 100\% branching fraction to top plus neutralino. The decay is carried out using phase space alone, so the decay products are completely unpolarized. This setup is exactly the CMS simplified model T2tt~\cite{Chatrchyan:2013sza}.
\item Right-handed stop production followed by decay to a bino-like LSP plus a top quark. In practice we take the exact setup for Scenario I but replace swap the roles of $\mu$ and $M_1$. This spectrum is close to the default signal model used by ATLAS\@. As the handedness of the stop and the identity of the LSP are fixed, the polarization of the emerging top quark is also fixed. 
\end{enumerate}
 By comparing our Scenarios with the stop signal models usually used, we can see how the Higgsino-like nature of the LSP and the hierarchy of the third generation squarks effects what regions of parameter space are allowed. The comparisons also give some indication of how well our simple analysis matches the full ATLAS/CMS results.

We can understand the strength of the bounds on our Scenarios by looking at the branching ratios and final states of our spectra. As we have decoupled the gauginos in all of our setups, the decays of the stops and sbottoms are governed entirely by the Yukawa couplings. For example, in Scenario I, all decays come from the top-Yukawa; up to kinematics, this yields a 50-50 split between decays to top quark plus neutralino and bottom quark plus chargino\footnote{The neutralino branching fraction is further split: roughly 50\% to $\neut_1$ and 50\% to $\neut_2$. However, this distinction does not make affect our analysis, since the two states have essentially the same mass}. Due to the degeneracy of the chargino-neutralino sector, chargino decay products are all extremely soft. In particular, the leptons from a chargino decay are far too soft to trigger the analysis requirements for the stop analysis, thus the only source of leptons is from the stops that decay to a leptonically decaying top quark. Additionally,
mixed decays $\tilde{t}_1\tilde{t}_1^* \ra t(\ra \ell \nu b) + \neut + b + \charpm_1$ may have a hard lepton, but they typically have fewer jets than required for a stop analysis. Therefore, only the fraction of events where both the stop and antistop decay to top + neutralino have a high probability of passing the analysis requirements.  As an final suppression, because the light stop in Scenario I is (almost) entirely right handed, the top quarks it yields are left-handed.  Due to the V-A nature of the weak interaction, left-handed tops have a softer lepton spectrum, and thus the leptons that the stops decays do create are less likely to pass the analysis cuts~\cite{ATLAS-CONF-2012-166}. The combined effect of these suppression factors is that there is no bound from the CMS direct stop search on Scenario I\@.

Similar logic works to understand the bounds on Scenario II and III\@. In Scenario II, both the $b_L$ and $t_L$ are produced. Up to effects of $O(y_b\,\tan{\beta})$ and ignoring kinematics, the only decay channel possible is $\tilde t \ra t + \neut$ for the stop and $\tilde b \ra t + \charm_1$ for the sbottom. The stop therefore decays in exactly the same way as in the default scenario, so we expect the bound to be at least as strong as the T2tt bound (with the added effect that the top is always right-handed and thus the emitted lepton is harder than in the unpolarized case). The bound is actually stronger because the sbottom decays also contribute; the chargino in a sbottom decay is indistinguishable from a neutralino, so the final state from a sbottom decay is virtually identical to the stop case. The only place the bound on Scenario II may weaken is close to or below the $t + \neut$ threshold, where $\tilde t \rightarrow b +\charp_1$ decays become important. Finally, we expect an even stronger bound in Scenario III\@. In addition to the $\tilde b$ decay that contributes exactly as in Scenario III, there are now two stop states and both states will contribute to the stop search. These suspicions are confirmed in  Fig.~\ref{fig:cmsstops}. 

\begin{figure}[t!]
\centering
\includegraphics[width=0.48\textwidth]{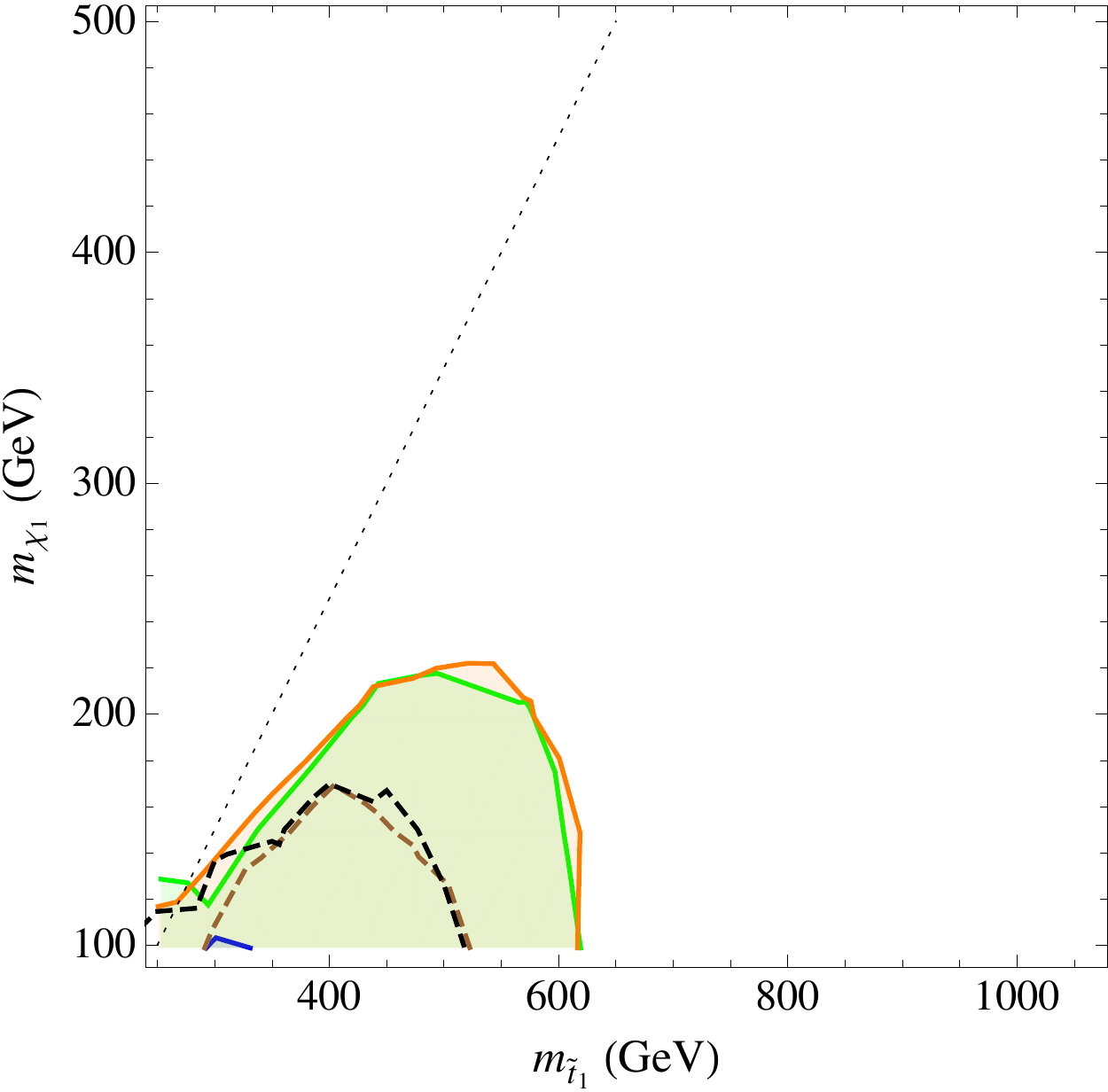}
\caption{Limits on the various 3rd generation scenarios coming from the 
ATLAS direct stop search (semileptonic channel). Contours are the same as in Fig.~\ref{fig:cmsstops}.}
\label{fig:atlstops}
\end{figure}

Moving to the ATLAS semi-leptonic stop search, we find similar results, shown in Fig.~\ref{fig:atlstops}. This is not surprising as the search criteria are very similar to the CMS stop search -- a single hard lepton and four or more hard jets. The biggest difference between the ATLAS and CMS semi-leptonic stop searches is that ATLAS requires a ``hadronic top candidate'' -- a three-jet subsystem with mass between $130\,\gev$ and $205\,\gev$ -- in all events. This requirement, along with slight changes in the analysis variables (see Appendix \ref{app:search}) lead to different  limits, but the qualitative message is the same as in the previous case: scenarios with $m_{\tilde{t}_R} \ll m_{\tilde{t}_L}, m_{\tilde{b}_L}$ are not bounded by these searches because the stops decay preferentially to $b + \charpm_1$ and therefore lack sufficient hard leptons and jet multiplicity, while scenarios with light $\tilde t_L, \tilde b_L$ are bounded tighter than the benchmark $\tilde t \rightarrow t + \neut$ scenario because both the stop and the sbottom decays contain top quarks\footnote{Comparing our bounds for $\tilde t_1\,\rightarrow t\,\neut$ (T2tt model) with the exclusions from ATLAS, we see a discrepancy -- our bounds are weaker by $O(100)\,\gev$. The fact that the DELPHES-based bound is different from the quoted number is not surprising, but the discrepancy is somewhat larger than expected. ATLAS has provided a cut-flow, at least for some benchmark $(m_{\tilde t_1}, m_{\neut_1})$ points, which allows us to pinpoint the difference to the $m_{jjj}$ cut (relative efficiencies of cuts either before or after this cut match to $O(10\%)$). We suspect the reason the $m_{jjj}$ cut is discrepant is that the jet-energy resolution in DELPHES is overly optimistic. If the jets retain too much of their energy, then the $m_{jjj}$ distribution will be shifted to higher values (relative to the full detector) and lost once the cut $m_{jjj} < 205\,\gev$ is imposed. If we increase the upper $m_{jjj}$ cut by $\sim 50\,\gev$, the signal efficiency at the benchmark point agrees better with the quoted value, however this artificial shift will have uncontrollable implications in the rest of the $(m_{\tilde t_1}, m_{\neut_1})$ efficiency plane. Therefore, we stay with the quoted cuts and emphasize that the relative bounds between models are the most relevant. The strong sensitivity of the bounds to $m_{jjj}$ also serves as a warning to the experiments since $m_{jjj}$ is susceptible to effects from ISR, the underlying event, and pileup. }.

The final direct-stop analysis we explore is an all-hadronic search performed by ATLAS using 20.5\, $\fbinv$ of data\footnote{ATLAS has performed a stop search in the dilepton final state~\cite{ATLAS-CONF-2012-167} assuming $BR[\tilde t_1 \ra b\,\chi^{\pm}] \sim 100\%$ and using a variety of  chargino-neutralino mass splittings (though none consistent with $\mu \ll M_1, M_2$). As the search requires two leptons, the same issues raised for Scenario I will be present and we expect no bound. For Scenarios II and III, we expect stronger limits from the semi-leptonic search since the decay $\tilde t_1\ra t\,\neut_0$ is dominant. For these reasons we do not explore the limits from the dileptonic searches on Natural Supersymmetry.}. Unlike the previous stop analysis, no leptons are involved. Instead, stops are searched for in events with multiple hard jets ($6$ or more), at least two $b$-jets, and substantial missing energy. To suppress multi-jet QCD backgrounds, the jets in the event are required to form two top-candidates -- three-jet subsystems with invariant mass between $80\,\gev - 270\,\gev$. When interpreted in terms of the $\tilde t_1 \rightarrow t + \neut$ benchmark scenario, ATLAS finds the strongest stop bound to date, nearly $700\,\gev$ for massless neutralino.

Applying these analyses to our three Scenarios, the bounds we find are shown below in Fig.~\ref{fig:atl_hadstops}. The trend of these bounds is similar to what we found in the previous stop searches, though the reasoning is slightly different. The bounds on Scenario II and III are nearly identical and rule out stops below $~750\,\gev$ for $\mu = 100\,\gev$. There is no significant bound on Scenario I due to the high fraction of decays to $b + \charpm_1$; stop decays to bottom quarks do not contain enough hadronic activity to efficiently pass the jet multiplicity cuts in this analysis.

\begin{figure}[t!]
\centering
\includegraphics[width=0.48\textwidth]{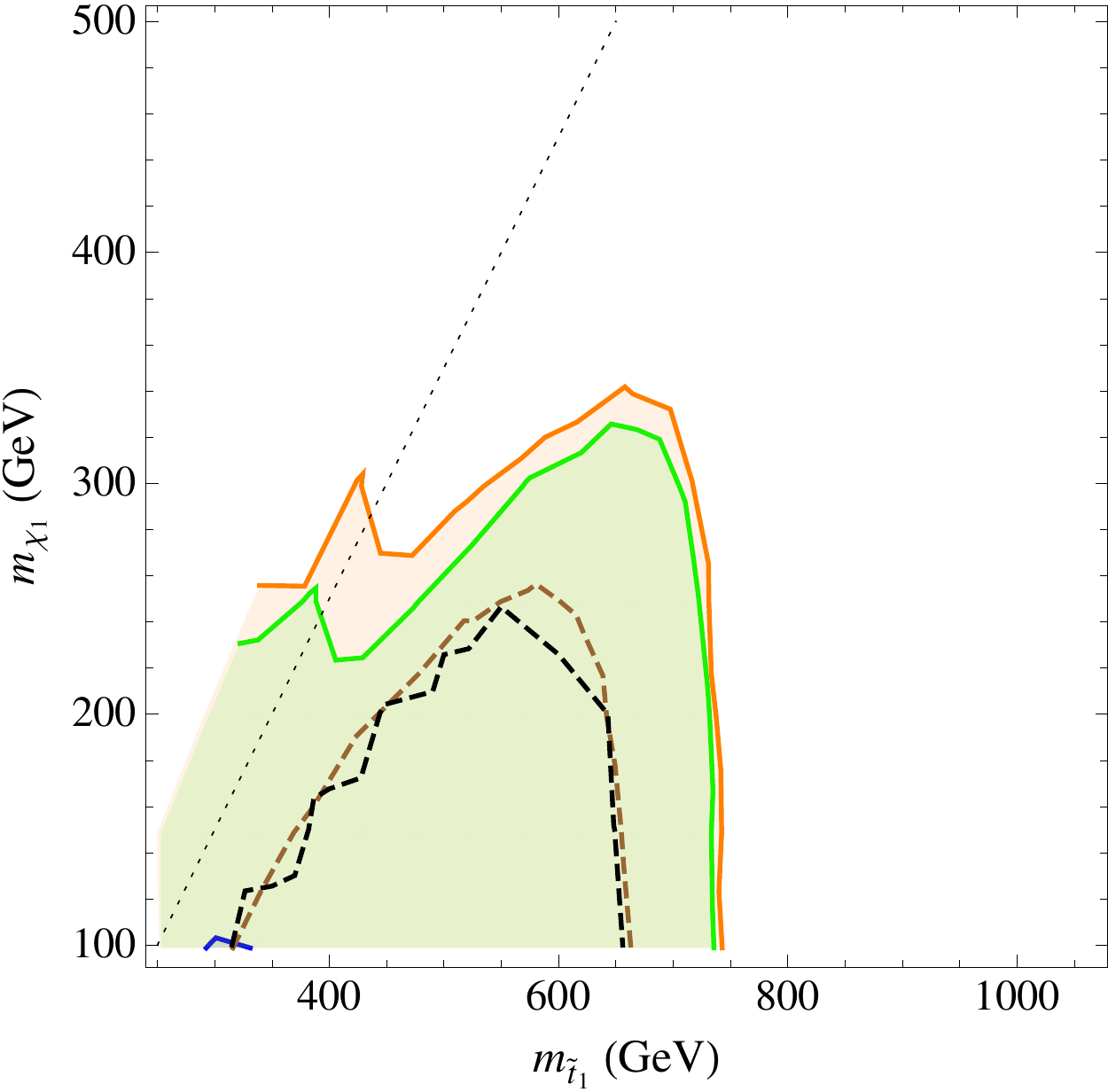}
\caption{Limits on the various 3rd generation scenarios coming from the 
ATLAS direct stop search (all-hadronic channel). Contours are the same as in Fig.~\ref{fig:cmsstops}.}
\label{fig:atl_hadstops}
\end{figure}

Summarizing the direct stop searches, scenarios with degenerate, light Higgsinos and $\tilde t_R \ll \tilde t_L, \tilde b_L$ are very weakly bounded, while the bounds on scenarios with light $\tilde t_L, \tilde b_L$ are quite tight, typically $100\,\gev$ stronger than the bounds on the benchmark $\tilde t_1 \rightarrow t +\neut$ setup. Because the direct stop searches are so insensitive to light $\tilde t_R$ (with light $\mu$), the bounds on Scenario II ($\tilde t_L, \tilde b_L$ and $\tilde t_R$ all light) and Scenario III (only $t_L, b_L$ light) are nearly identical. However, before we can draw any firm conclusions on Natural Supersymmetry, we must also consider the CMS and ATLAS searches tailored towards the detection of sbottoms.

\subsection{Direct sbottom searches}
\label{sec:sbot}

In Natural Supersymmetry, the stops can decay into $b + \charpm_1$,
and thus dedicated searches for $b$-quarks plus missing energy are 
vital to our analysis.  In addition, in both Scenarios II and III,
$\tilde{b}_L$ is present in the spectrum with 
$m_{\tilde{b}_L} \simeq m_{\tilde{t}_L}$ determined by $m_{Q_3}$. 
In this section we use the ATLAS and CMS searches that target 
direct sbottoms~\cite{Chatrchyan:2013lya,atlas_sbottoms}, since these 
studies focus on $b$-jets and missing energy and are therefore independent 
of the mass splittings in the chargino/neutralino sector.

To isolate signal-rich regions from background, ATLAS/CMS sbottom searches require multiple high-$p_T$ jets along with one or more flavor tags. Events with leptons are vetoed as a way to remove some $\bar t t$ background (the leptonic events are retained as control samples).  More elaborate cuts are applied to further enhance the signal depending on the collaboration and the target signal mass. The default signal we will compare to is pair-production of sbottoms that decay solely to $b$ quarks plus neutralino, a $\bar b b + \slashchar{E}_T$ final state. Since it is identical to the CMS default signal model, we will refer to the default as T2bb as they do.

To bound $\bar b b + \slashchar{E}_T$ signals, CMS~\cite{Chatrchyan:2013lya} retains events with 2-3 jets and 1 or 2 $b$-tags. The visible objects in the event are partitioned into two ``mega-jets''. The degree to which these mega-jets balance each other, described with the $\alpha_T$ variable~\cite{Randall:2008rw, Chatrchyan:2011zy}, as well as the net $H_T$ are used to further isolate the signal from background. The bounds from this analysis on the CMS default model and on our three scenarios are shown below in Fig.~\ref{fig:cmssbots} 

\begin{figure}[t!]
\centering
\includegraphics[width=0.48\textwidth]{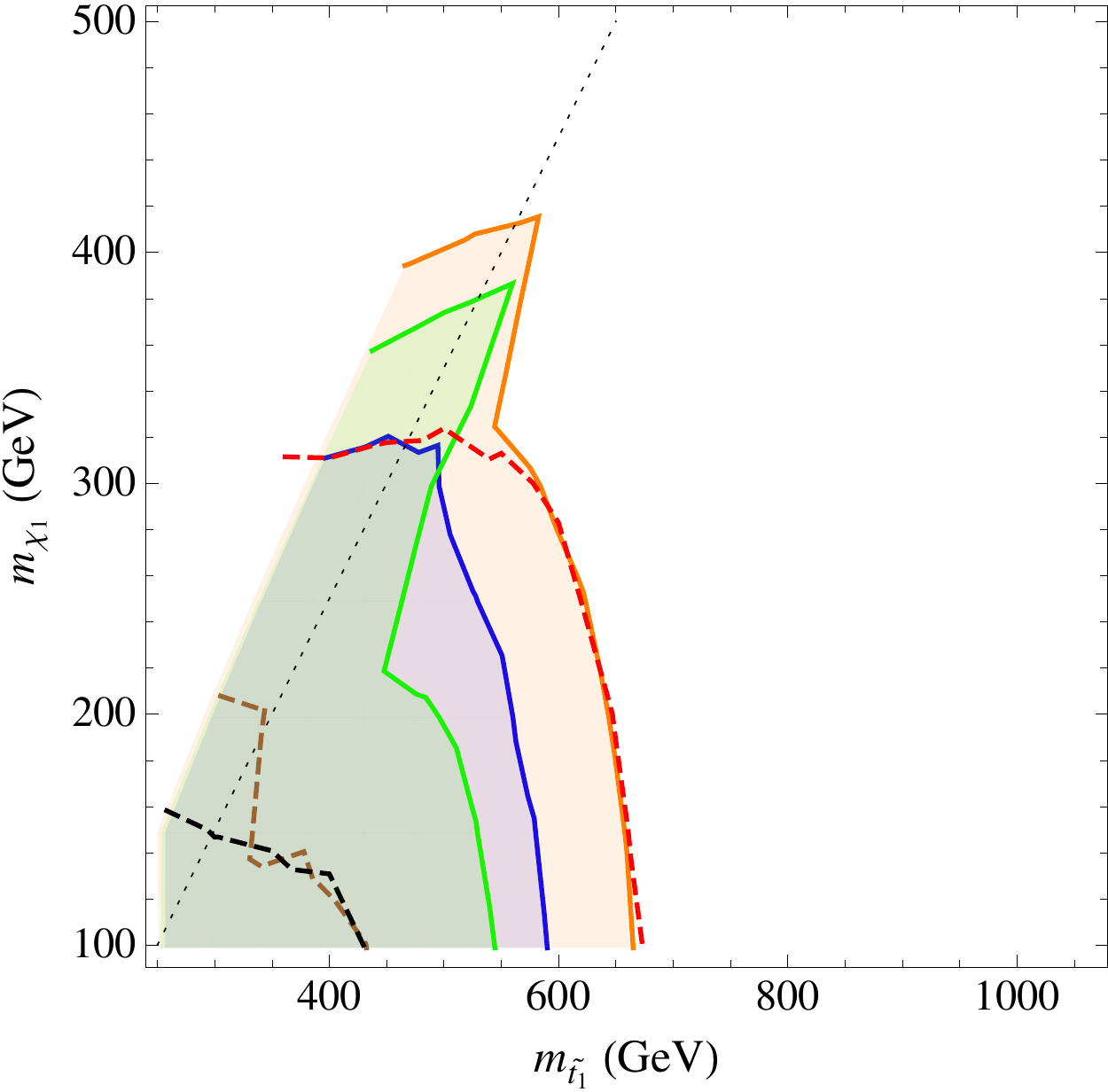}
\caption{Limits on our 3rd generation scenarios from the CMS direct sbottom search. Contours are the same as in Fig.~\ref{fig:cmsstops}, however there is now a bound, indicated in blue, on Scenario I\@. We have also added the bound (red dashed line) derived from applying this analysis to the T2bb simplified model, direct production of right-handed sbottoms with 100\% branching fraction to a bottom quark and a neutralino. The remaining contours are the same as in Fig.~\ref{fig:cmsstops}.}
\label{fig:cmssbots}
\end{figure}

The first thing to notice is that the sbottom search places a strong bound on Scenario I -- roughly $m_{\tilde t_1} > 600\,\gev$ for $m_{\neut_1} \sim \mu \sim 100\,\gev$ and decreasing slightly as $m_{\neut_1}$ is raised.  

The bounds on Scenario I are weaker than the bounds on the T2bb scenario. This is because Scenario I yields more leptons -- coming, as before, from stop decays to leptonic tops -- so events from Scenario I are more likely to be vetoed. Also, the average number of jets is higher, pushing the signal into jet bins not considered in the sbottom analysis. The same two effects also explain the difference in bounds between Scenarios I and II\@. In Scenario II, provided $m_{\tilde t_1} \gg m_t + m_{\neut_1}$, both stop and sbottom decays result in top quarks. The only di-top quark events that cleanly mock the $\bar b b + \slashchar{E}_T$ signal are fully leptonic events where both leptons are lost or lie outside the tracking volume. In all other events there is either a lepton or a larger jet multiplicity and the event is either vetoed or populates a region not usually considered as signal. The caveat to this argument is when $m_{\tilde t_1} \lesssim m_t + m_{\neut_1}$. In this region, kinematics suppresses the $\tilde t_1 \rightarrow t + \neut$ mode and the (otherwise Yukawa suppressed) $\tilde t_1 \rightarrow b + \charpm_1$ mode becomes important. Decays to $b + \charpm_1$ are efficiently selected by the CMS search, explaining why the bound on Scenario II gets stronger the closer the stop mass gets to $m_t + m_{\neut_1}$. The bound in the threshold region of Scenario II is actually stronger than in Scenario I since both $\tilde{t}_L$ and $\tilde{b}_L$ are produced and both decay to $b + \tilde{\chi}$ when $m_{\tilde{t}_1} \sim m_{\tilde{b}_1} \lesssim m_t + m_{\neut}$. As expected, the bound on Scenario III is the strongest and resembles the sum of the bounds on Scenario I and II\@.

The ATLAS direct sbottom~\cite{atlas_sbottoms} search targets the same final state, $\bar b b + \slashchar{E}_T$ as the CMS search. However, the ATLAS search is more optimized to the topology with exactly two bottom jets, missing energy, and little other hadronic activity. A third hard jet is vetoed in the majority of the analysis channels, and no channel tolerates 4 or more jets. As a result, the ATLAS sbottom search is less flexible and not as well suited to events that contain top quarks. The bounds from the ATLAS sbottom search cast in term of our scenarios and the benchmark T2bb model are shown below in Fig.~\ref{fig:atlsbots}\footnote{In the ATLAS analysis~\cite{ATLAS-CONF-2013-001} the sbottom search technique was used to constrain stop production, exactly as we are advocating here. However in that analysis, $BR(\tilde{t}_t \ra b\,\charpm_1) = 1$ was assumed, so bounds presented there do not constrain scenarios with a light Higgsino, a key ingredient in Natural Supersymmetry.}.
\begin{figure}[t!]
\centering
\includegraphics[width=0.48\textwidth]{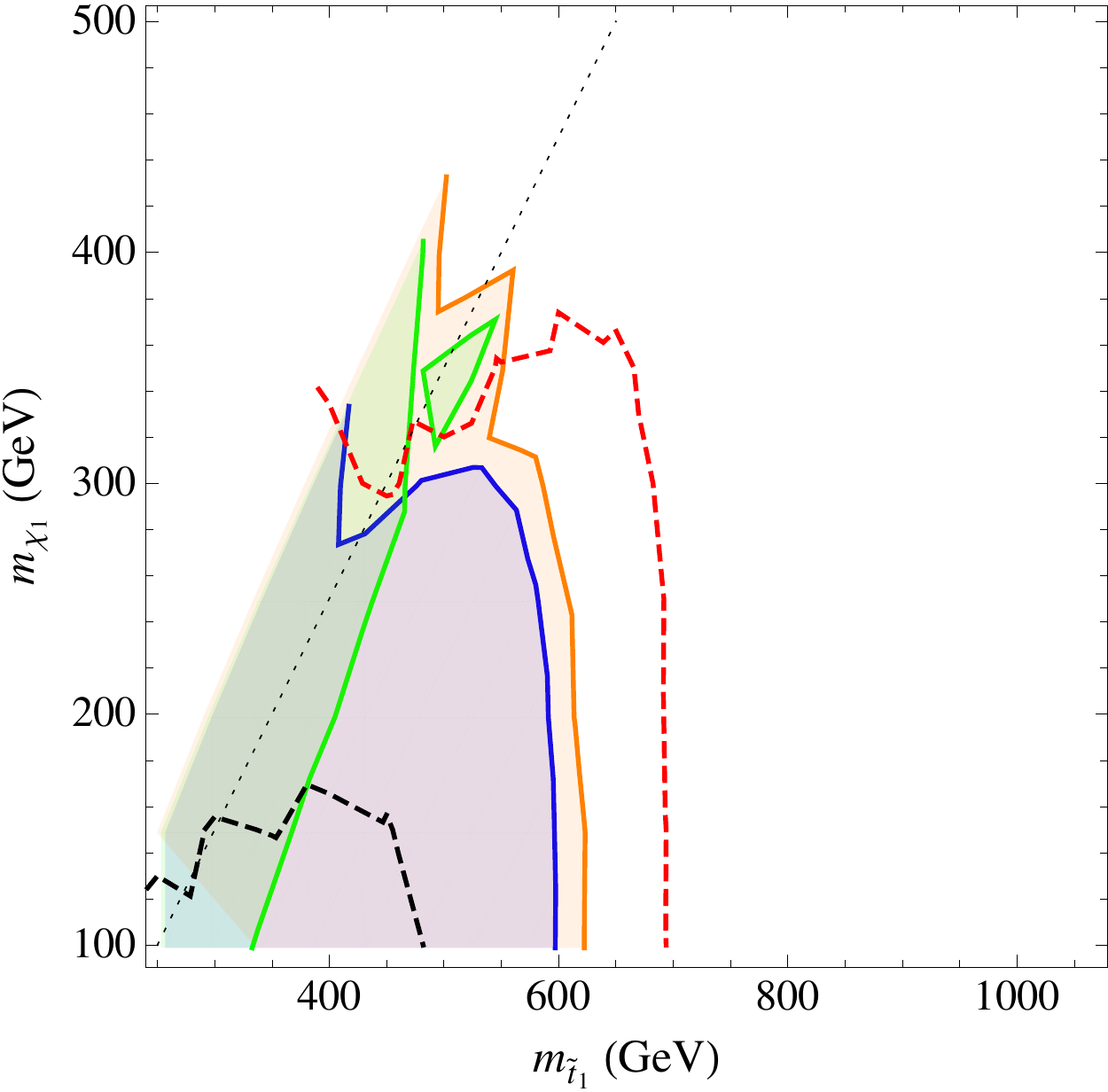}
\caption{Limits on our 3rd generation scenarios from the ATLAS direct sbottom search. Contours are the same as in Fig.~\ref{fig:cmssbots}.}
\end{figure}


\subsection{Combined Bounds}
\label{sec:combinedboundds}

Combining the three stop searches and two sbottom searches by taking the strongest limit at a given $(m_{\tilde t_1}, m_{\neut_1})$ point, we get the net excluded region for the three Scenarios. The excluded regions are displayed below in Fig.~\ref{fig:combined} along with the analogous regions for the default spectra.

\begin{figure}[t!]
\centering
\includegraphics[width=0.48\textwidth]{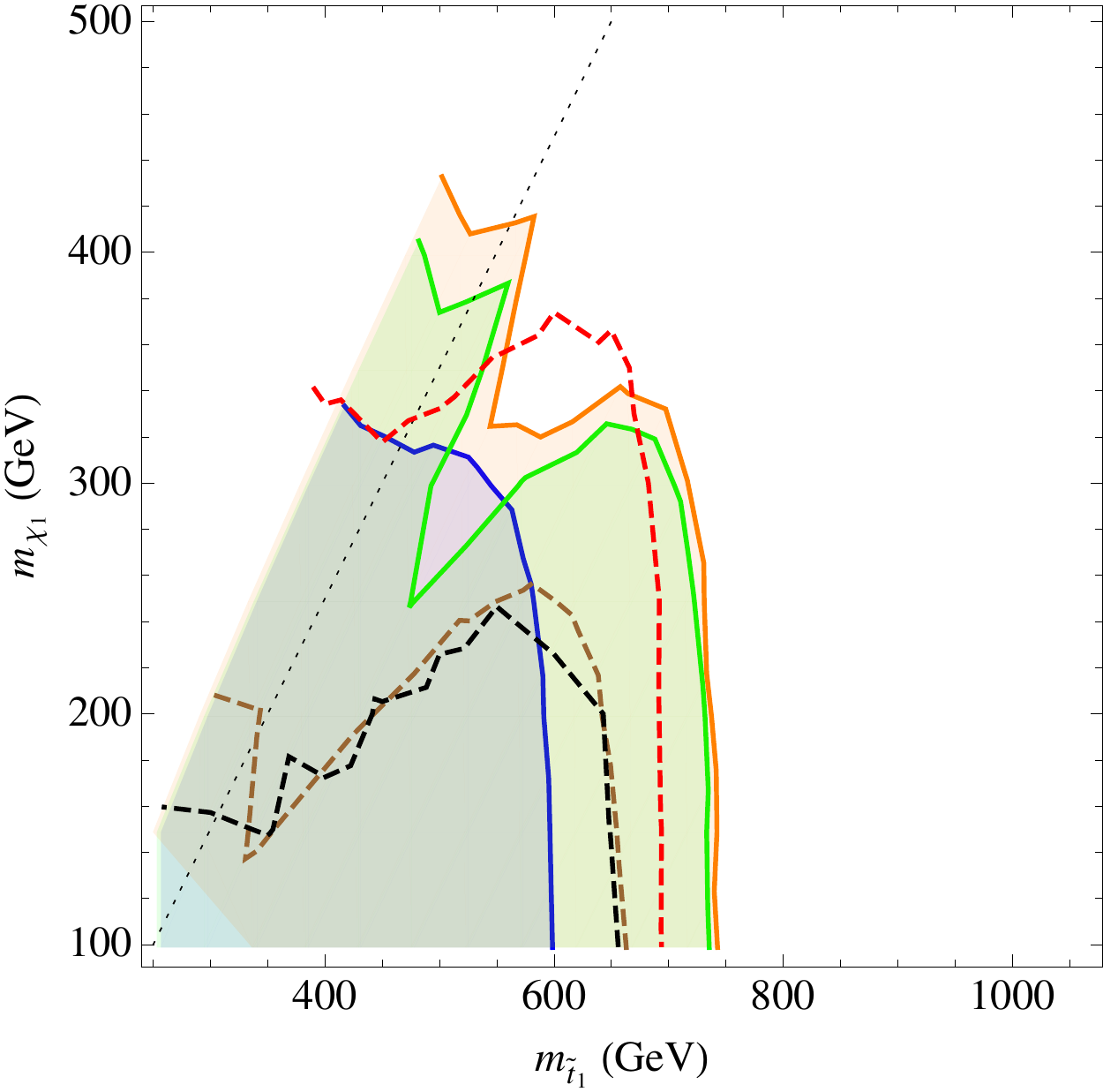}
\caption{Limits on our 3rd generation scenarios from combining all CMS and ATLAS sbottom/stop searches search. Contours are the same as in Fig.~\ref{fig:cmssbots}.}
\label{fig:combined}
\end{figure}


\section{Implications for the Higgs sector}
\label{sec:precisionhiggs}

In this section we study the implications of Scenario I, II and III on 
the supersymmetric Higgs sector. 
In supersymmetry the additional charged and colored degrees of freedom 
can significantly modify the production cross section and branching ratios 
of the lightest (standard model-like) Higgs 
boson~\cite{Shifman:1979eb,Kniehl:1995tn,Kane:1995ek,Kileng:1995pm,Spira:1995rr,Dawson:1996xz,Djouadi:1996pb,Djouadi:1998az,Belanger:1999pv,Harlander:2004tp,Dermisek:2007fi,Bonciani:2007ex,Low:2009di,Harlander:2010wr}.
Given the recent discovery of a particle consistent with a Higgs boson at 
$m_h \simeq 125$~GeV \cite{Aad:2012tfa,Chatrchyan:2012ufa}, 
the modifications due to the additional charged and colored degrees of 
freedom have been extensively studied~\cite{Arbey:2011ab,Heinemeyer:2011aa,Draper:2011aa,Carena:2011aa,Ellwanger:2011aa,Blum:2012ii,Buckley:2012em,Espinosa:2012in,Carena:2012gp,Carena:2013qia,Carena:2013iba}.

First let us consider the Higgs boson branching ratios.  When the stop 
contributions are included, the modification to the decay rate of the 
Higgs boson into gluons is given by~\cite{Djouadi:2005gj,Dermisek:2007fi}
\begin{eqnarray}
\frac{\Gamma_{ggh}^{\rm MSSM}}{\Gamma_{ggh}^{\rm SM}} &\simeq& 
\left| 1 + \frac{1}{A_{1/2}(\tau_t)} \sum_{i=1}^2 \frac{g_{h\tilde{t}_i \tilde{t}_i}}{m_{\tilde{t}_i}^2} A_0 (\tau_{\tilde{t}_i})  \right. \nonumber \\ 
& & \left. +
\frac{1}{A_{1/2}(\tau_t)} \sum_{i=1}^2 \frac{g_{h\tilde{b}_i \tilde{b}_i}}{m_{\tilde{b}_i}^2} A_0 (\tau_{\tilde{b}_i}) \right|^2 
\end{eqnarray}
where $A_{1/2} (A_0)$ are the standard fermion (scalar) loop functions
(e.g.~\cite{Djouadi:2005gj}), and $\tau_i = m_h^2/4m_i^2$.
Here $m_{\tilde{f}_2} \geq m_{\tilde{f}_1}$, $\theta_f$ is the sfermion the mixing angle, and the couplings are given by 
\begin{equation}
 \frac{g_{h\tilde{f}_i \tilde{f}_i}}{m_{\tilde{f}_i}^2}  \simeq \frac{m_f^2}{m_{\tilde{f}_i}^2} + \frac{(-1)^i}{4} s_{2\theta_f}^2 \frac{m_{\tilde{f}_2}^2 - m_{\tilde{f}_1}^2 }{m_{\tilde{f}_i}^2}+ \mathcal{O}\left(\frac{M_Z^2}{m_{\tilde{f}_i}^2}\right) \, ,
\end{equation}
in the decoupling limit.
Hence in the limit of small mixing, $\theta_f \sim 0$, the squarks enhance the decay rate of the Higgs boson into gluons. Similarly, for sbottoms the contributions are typically small except in the large $\tan \beta$ regime where sbottom contributions will interfere destructively with the top contribution.

Light stops, sbottoms, staus and charginos also affect the decay of the Higgs boson into photons. Enhancing the decay rate of the Higgs to gluons due to light stops will lead to a suppressed decay rate of the Higgs into photons due to the stop  contribution destructively interfering with the $W$-boson contribution (the dominant standard model contribution), while a light sbottom has the opposite effect. Furthermore, depending on the sign of $\mu$, a light Higgsino close to the LEP bound~\cite{LEPbound}  can either enhance or suppress the photon rate~\cite{Huo:2012tw}. Expanding in terms of the $1/M_2$, where $M_2$ is the Wino mass parameter, we find the amplitude of the lightest chargino is
\begin{eqnarray}
|A_{\charpm}|  \approx \frac{2 M_W^2}{|M_2| m_{\neut_1}} |c_{\alpha +\beta}| A_{1/2}(\tau_{\neut_1})
\end{eqnarray}
where $M_W$ is the W-boson mass, $M_2$ is the Wino mass, $m_{\neut_1} \sim \mu$ and 
$\alpha$ is the mixing angle of the CP-even Higgs bosons. In the decoupling limit $c_{\alpha + \beta} \sim s_{2\beta}$, this contribution becomes suppressed for large $\tan \beta$~\cite{Huo:2012tw}. Therefore the
charged contributions are also relatively suppressed in an MSSM-like framework. 

\begin{figure*}
\begin{center}
\includegraphics[width=0.32\textwidth]{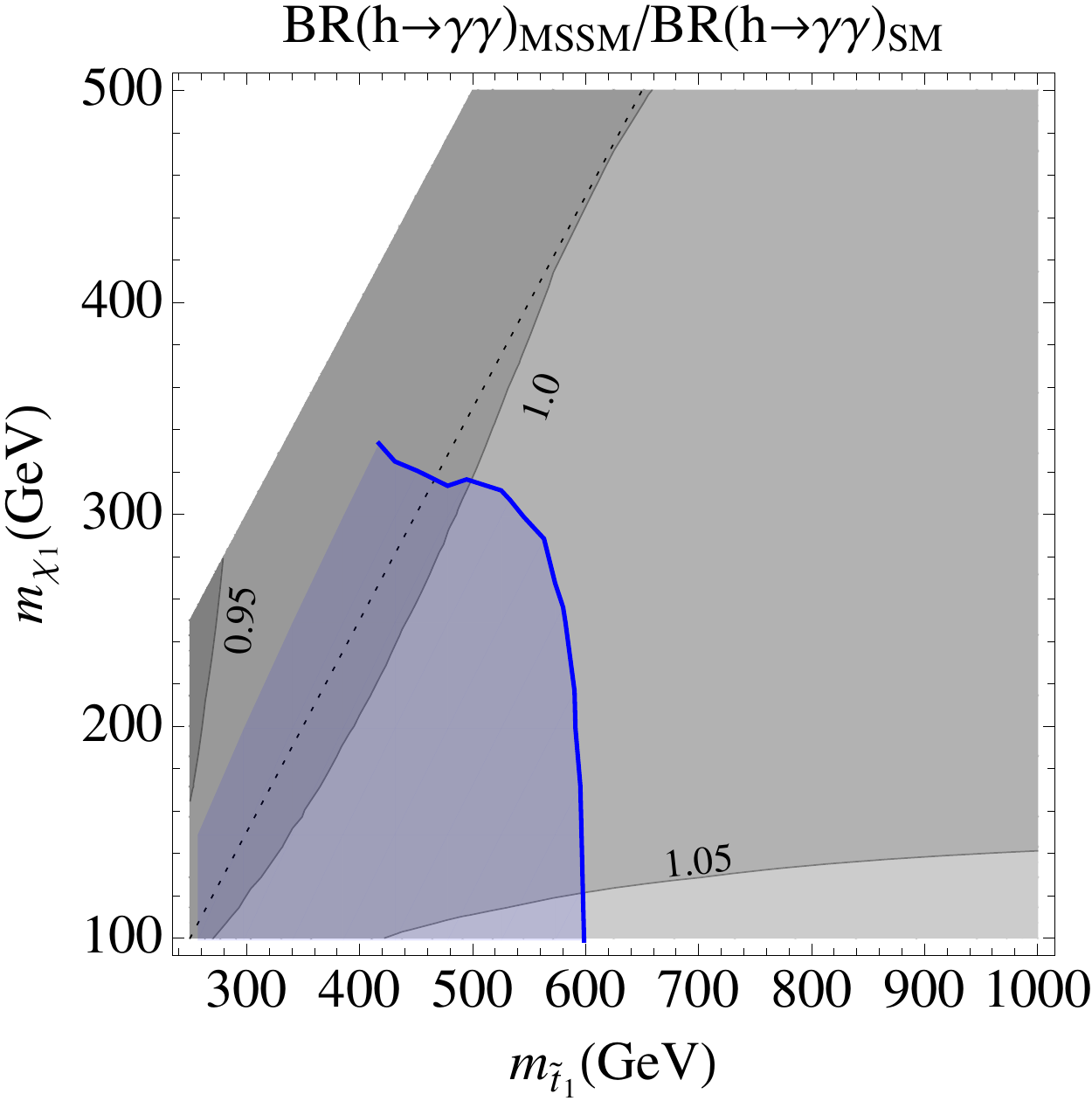}
\includegraphics[width=0.32\textwidth]{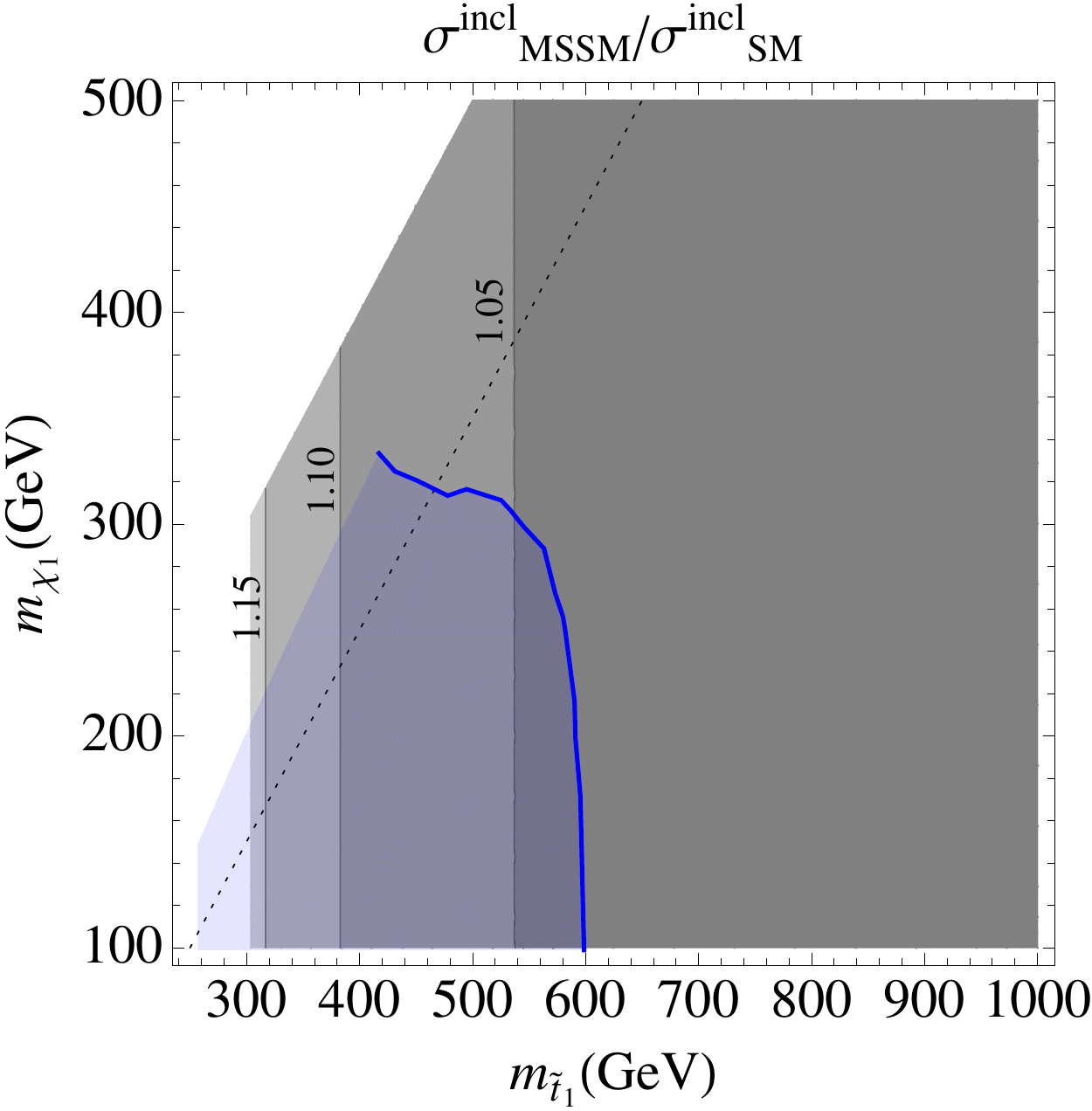}
\includegraphics[width=0.32\textwidth]{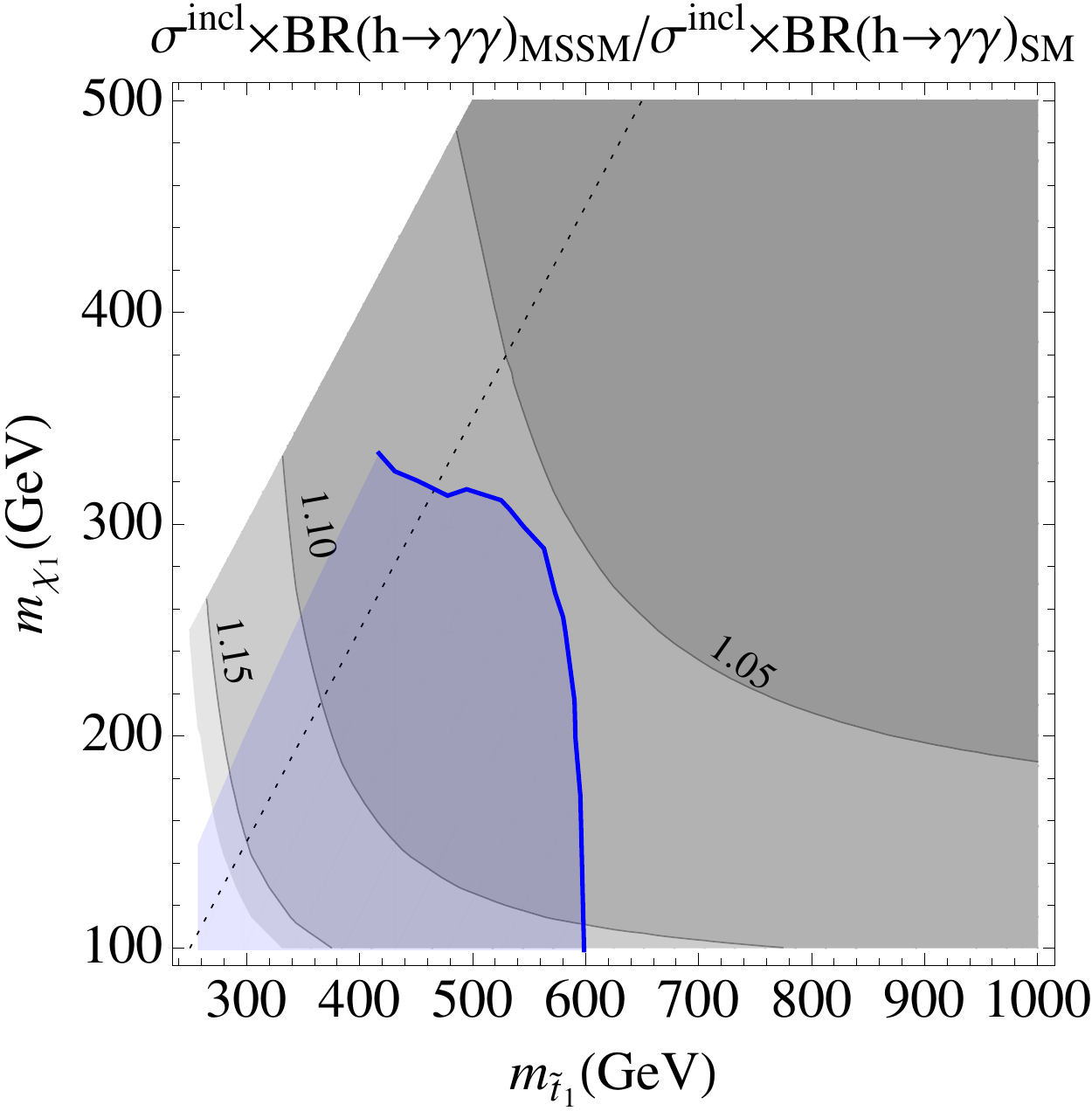}\\
\includegraphics[width=0.32\textwidth]{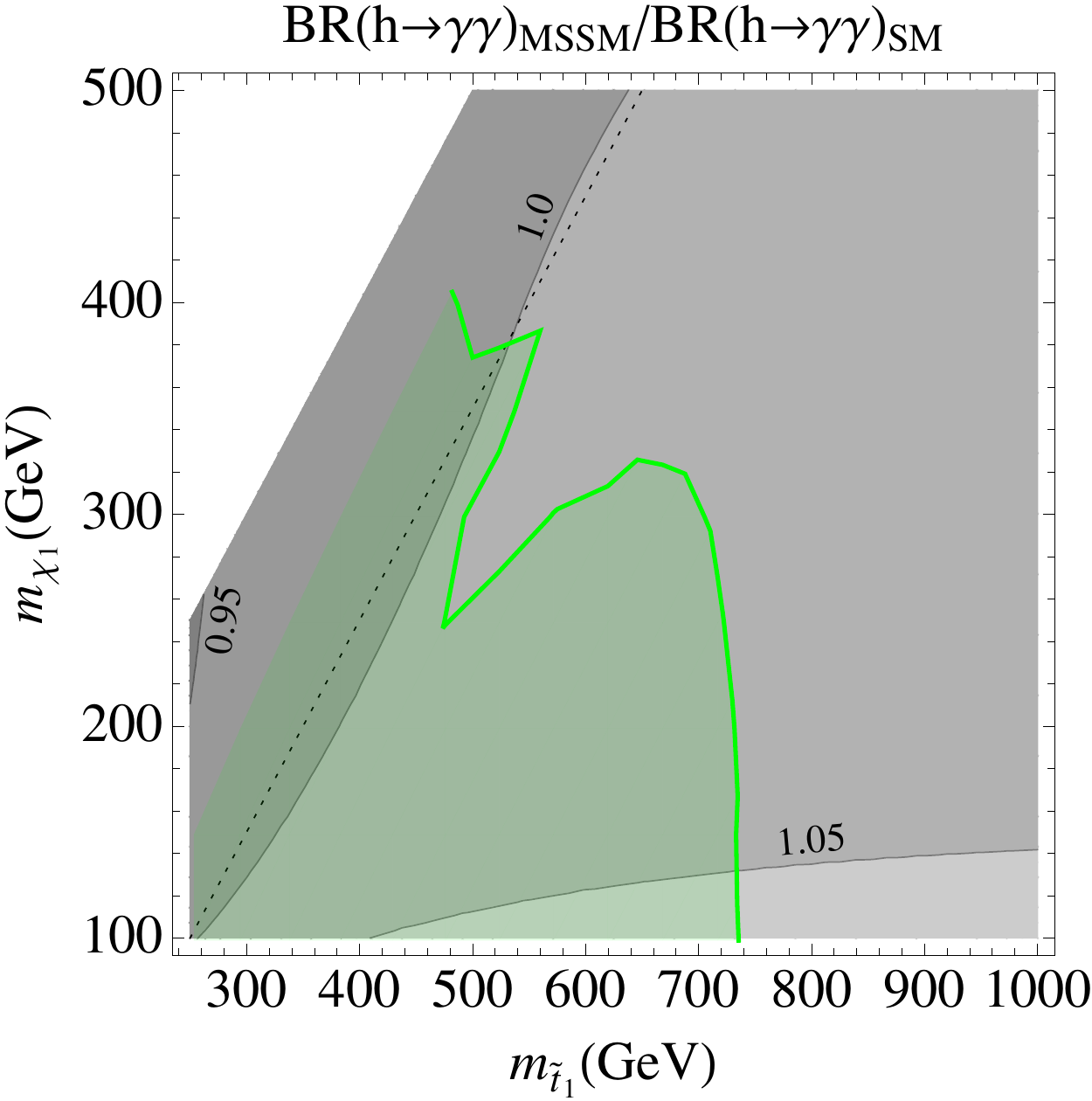}
\includegraphics[width=0.32\textwidth]{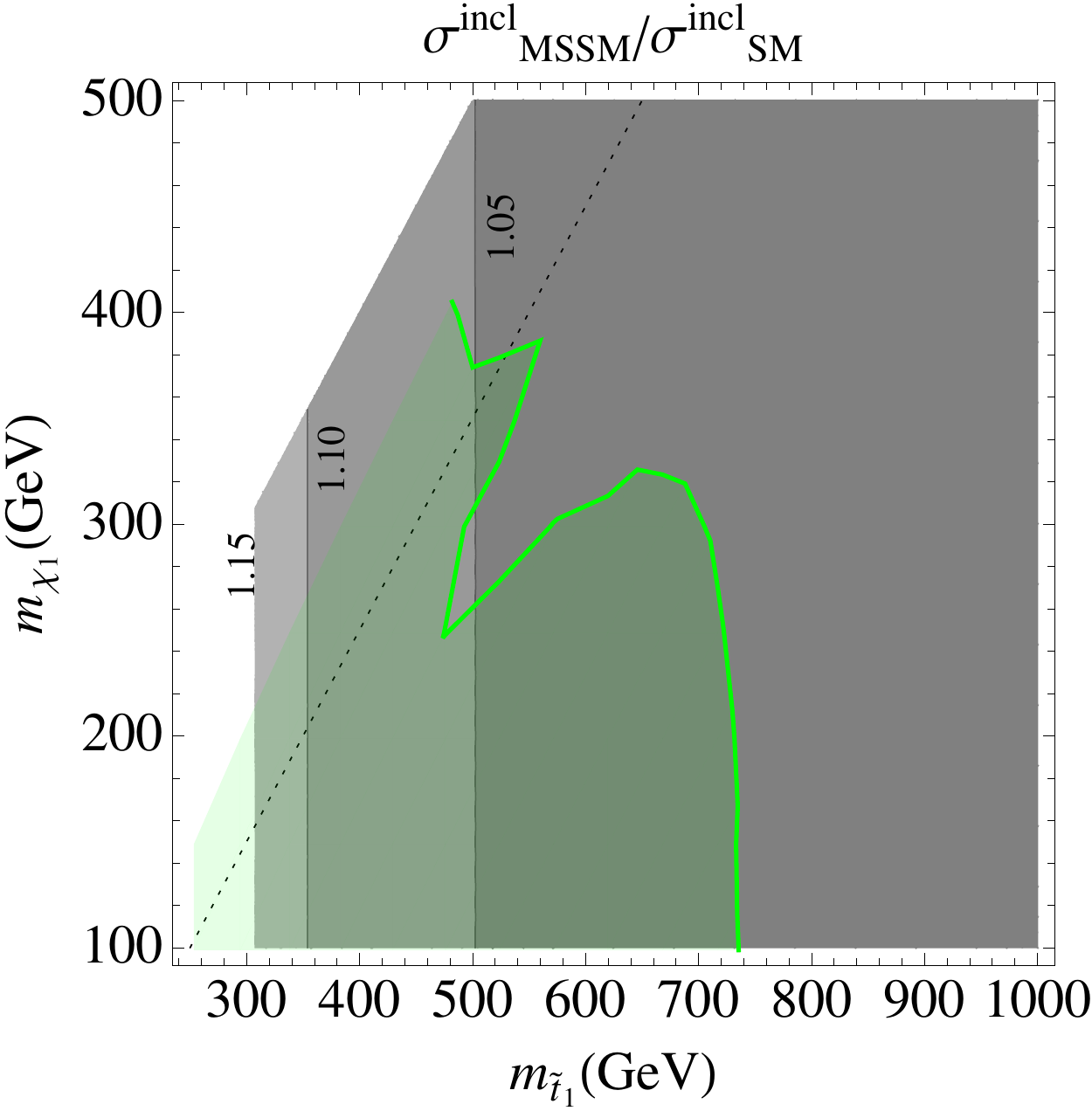}
\includegraphics[width=0.32\textwidth]{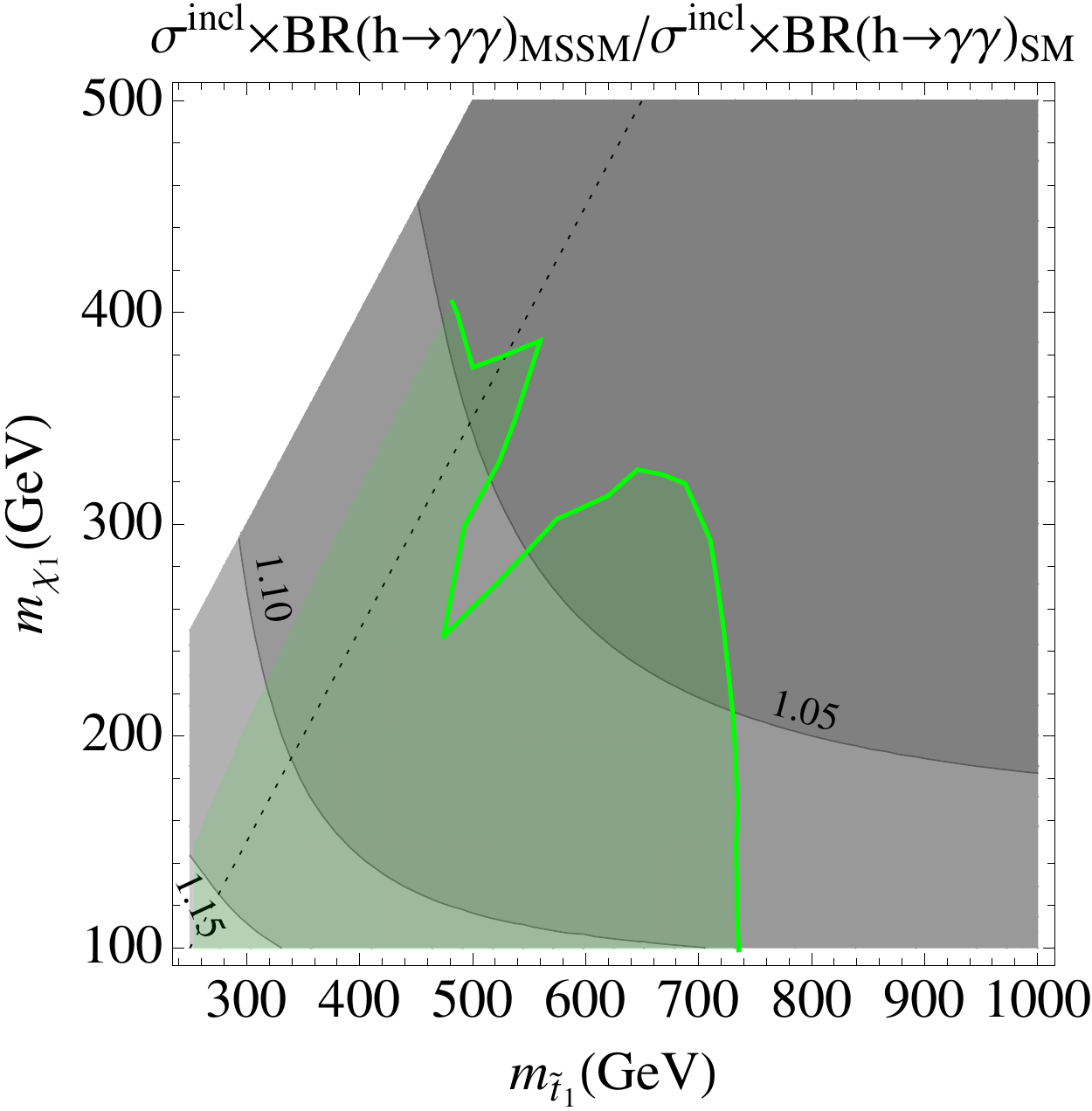}\\
\includegraphics[width=0.32\textwidth]{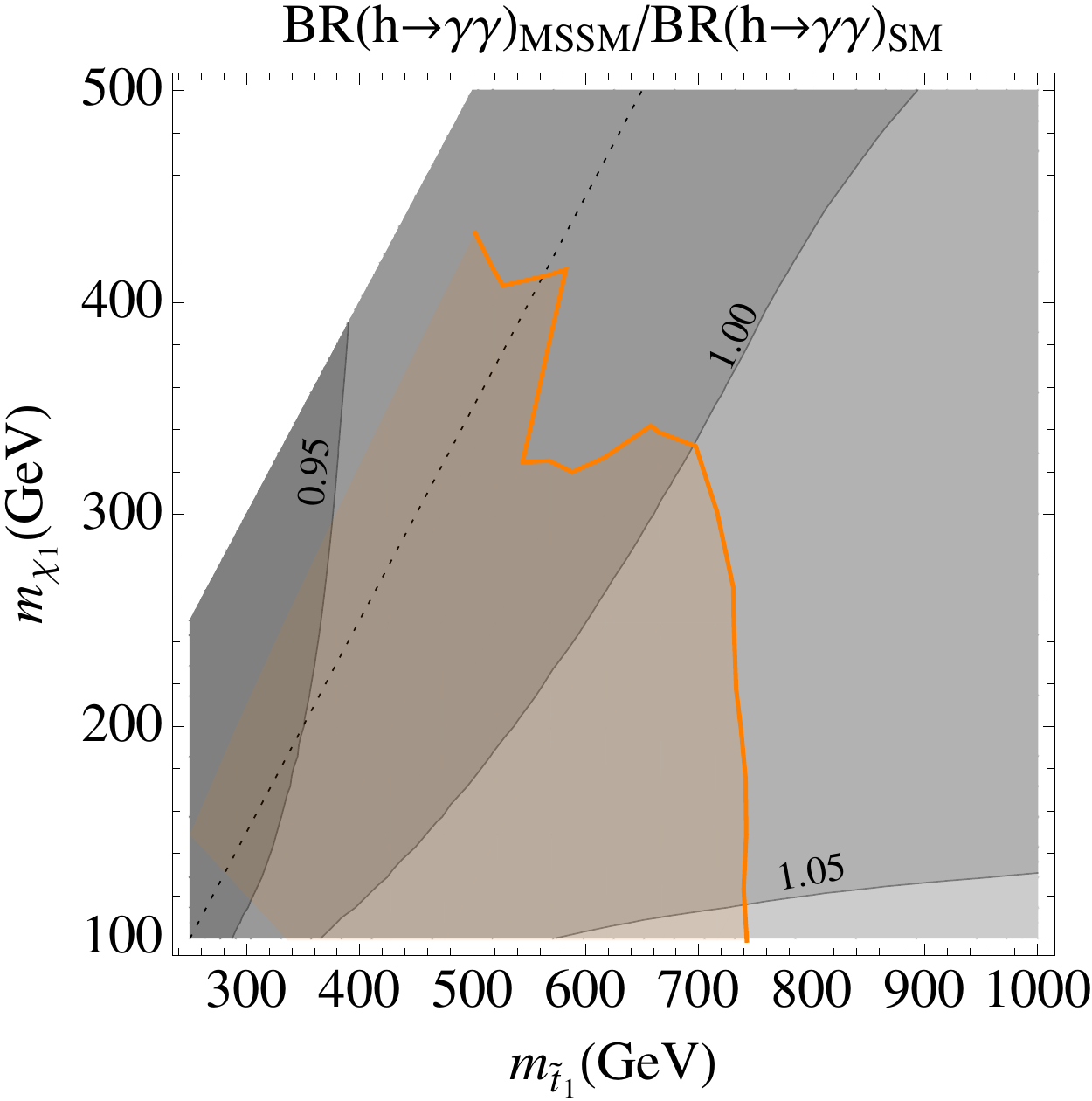}
\includegraphics[width=0.32\textwidth]{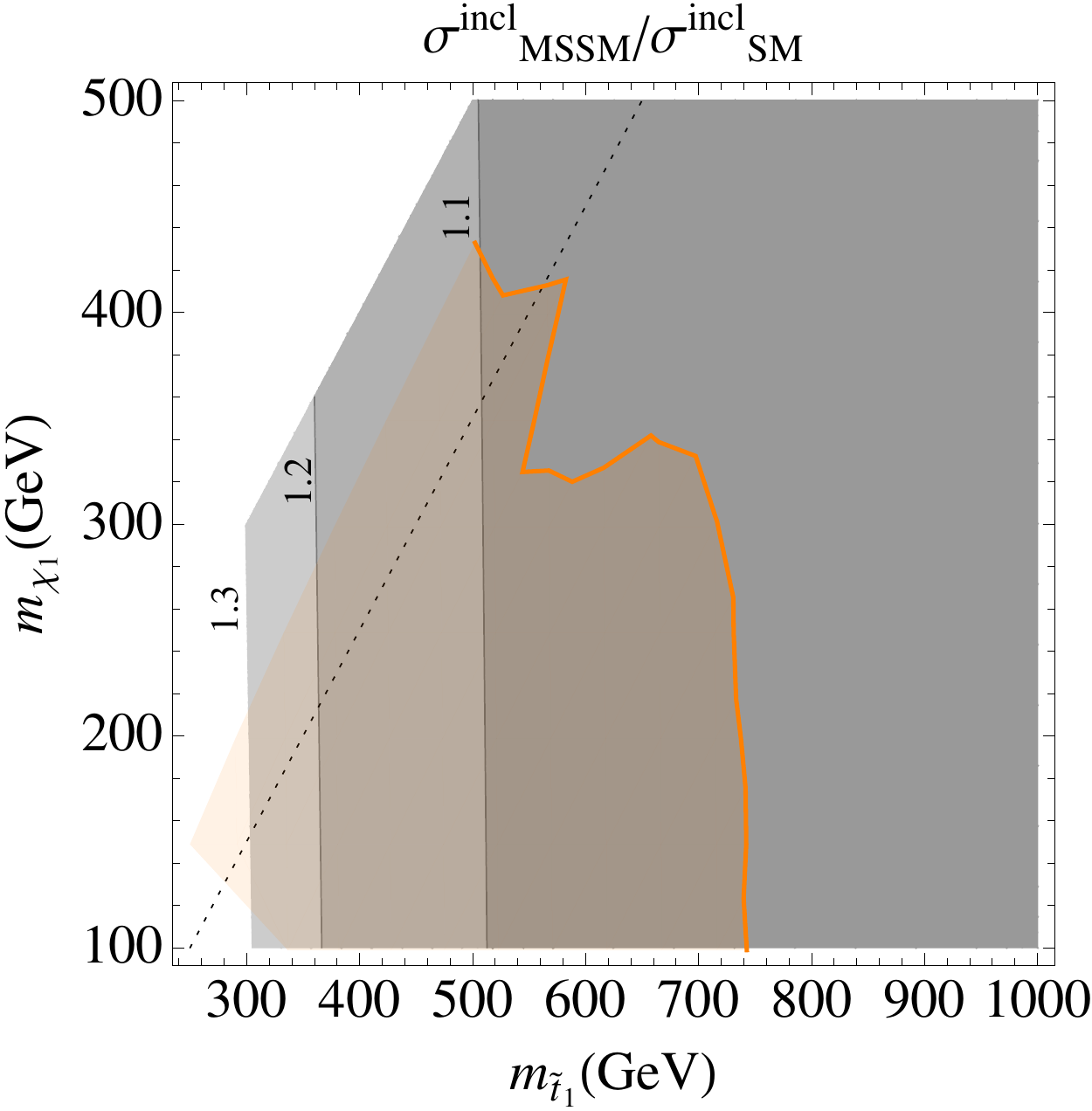}
\includegraphics[width=0.32\textwidth]{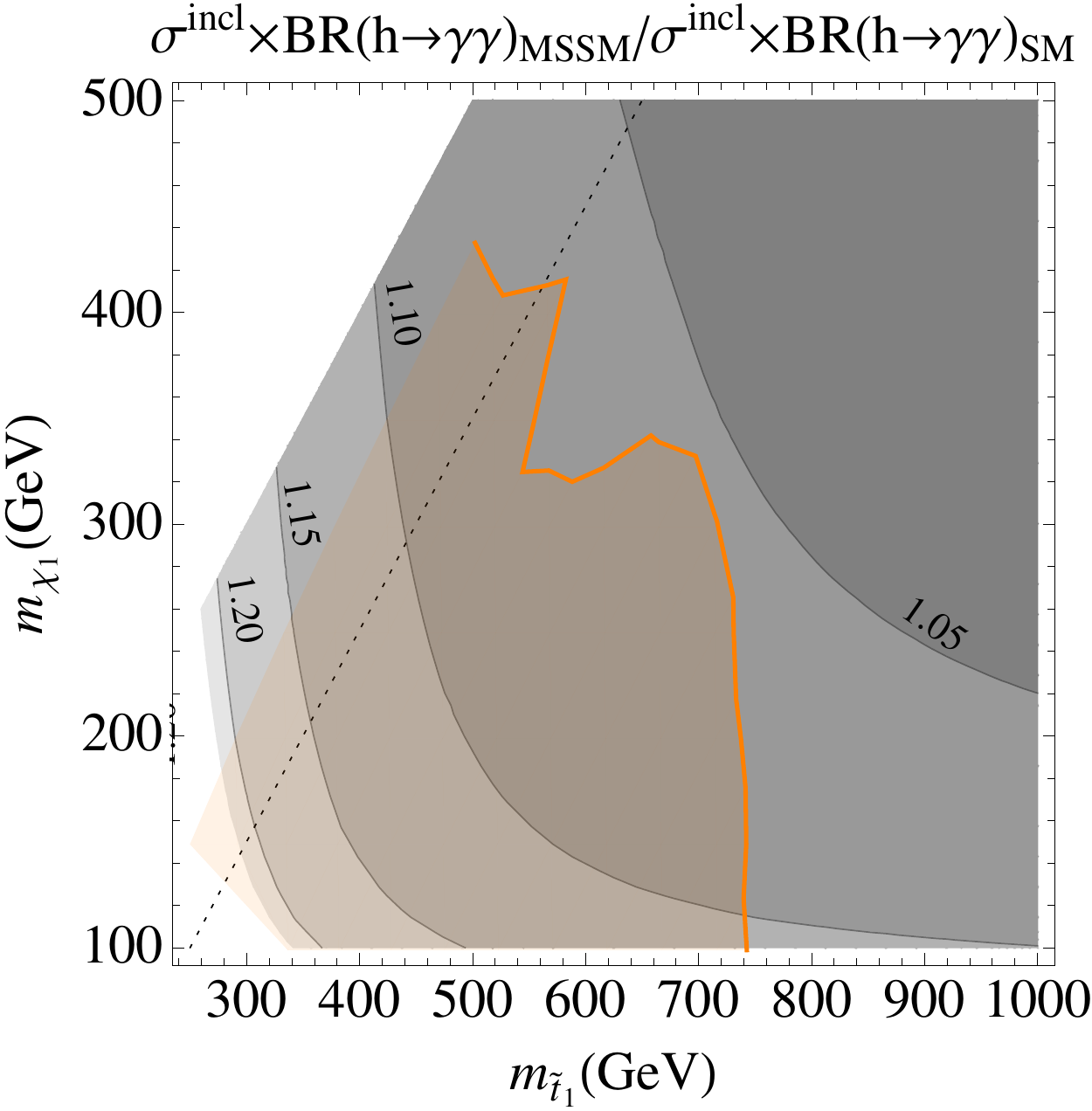}\
\end{center}
\caption{Modifications to Higgs production and branching ratios in the 
decoupling limit where $M_A \ra \infty, A_t = A_b = 0$ and $M_2 = 1$~TeV\@.
We have overlayed the direct search constraints found in Fig.~\ref{fig:combined}
(same coloring).  The top, middle, and lower set of figures correspond to 
Scenario I, II, and III\@.
The ``compressed wedge'' corresponds to the allowed region in the 
upper-left of each plot, where the mass difference between the squark 
and the Higgsinos is small.  The ``kinematic limit'' region 
corresponds to the allowed region to the far-right of each plot,
where the squark production cross section reaches the kinematic
limit of the LHC searches.}
\label{fig:mainf}
\end{figure*}

In Fig.~\ref{fig:mainf} we show the impact of the spectra in Scenario I, II and III on modifications to $BR(h \to\gamma \gamma)$, $\sigma_{incl}$, and $\sigma_{incl} \times BR(h \to \gamma \gamma)$.  This is the principle result of our paper. We have taken $\mu > 0$ and $M_A = \infty$, however the results are nearly identical (to within $\simeq 5\%$) when $M_A = 1000$~GeV\@.\footnote{For lower values of $m_A$, the increased mixing between the two CP-even Higgs bosons leads to a slight further suppression in the branching ratios.} Here we have assumed there is an additional contribution to the quartic coupling, raising the Higgs mass up to the experimentally measured value $m_h \simeq 125$~GeV, such as the NMSSM-like scenario described in Appendix~\ref{sec:NMSSM}. 

We see that the modifications to the inclusive production cross section (dominated by gluon fusion) are at the $10$-$30\%$ level in the ``compressed wedge'', while rather small $\lsim 5\%$ at the ``kinematic limit''.  The $BR(h \to \gamma \gamma)$ receives considerably smaller effects, between $-5\%$ to $+5\%$ across the parameter space of interest.  These deviations are not large enough to be directly constrained by the measurements from the LHC~\cite{LHChiggs} and Tevatron~\cite{Aaltonen:2013kxa}, however as we measure the Higgs production and branching ratios more precisely, we expect these deviations to be observable at the LHC\@.


\section{Discussion}

We have shown that Natural Supersymmetry, where third generation squarks
decay into Higgsino-like neutralinos and chargino, is significantly constrained
by existing LHC searches, summarized in Fig.~\ref{fig:combined}.  
When these constraints are overlayed on the modifications to the
lightest Higgs production and decay, shown in Fig.~\ref{fig:mainf}, 
we find distinctly different
implications for the remaining allowed regions identified as the 
``compressed wedge'' ($(m_{\tilde{q}} - |\mu|)/m_{\tilde{q}} \ll 1$)
and the ``kinematic limit'' ($m_{\tilde{q}} > 600$-$750$~GeV).
We found that the collider constraints arise from the totality of numerous searches 
at ATLAS and CMS that are separately sensitive, in varying degrees, to
squark production and decay through 
$\tilde{t} \ra t \neut$, $\tilde{t} \ra b \charp_1$, 
$\tilde{b} \ra b \neut$, and $\tilde{b} \ra t \charm_1$.
Our analysis incorporated simulations of the signal and detector response,
matching the experimental analyses as close as we could.  
Nevertheless, there is substantial room for improvement.
Much of the experimental searches were designed for only one
decay mode, or chose chargino/neutralino mass hierarchies 
that are not consistent with Natural Supersymmetry. 
We believe dedicated analyses, that take into account
the proper branching fractions and mass hierarchy, may well 
significantly improve the sensitivity.

Natural Supersymmetry implies the wino and bino are sufficiently
heavy that the Higgsino-like chargino and neutralino splittings
are very small, just several GeV\@.  While we focused our attention
on the stop/sbottom signals, the electroweakinos (Higgsinos) can be
directly produced at the LHC\@.  However, the narrow splittings
of the Higgsinos makes them extremely difficult to detect; 
the traditional search for electroweakinos is 
$pp \ra \neut_2 \charpm_1 \rightarrow 3\ell + \slashchar{E}_T$~\cite{trileptons},
where the leptons come from cascade decays 
$\neut_2 \ra Z\,\neut_1, \charpm_1 \rightarrow W^{\pm}\,\neut_1$. 
As the spectrum gets squeezed, the intermediate $W^{\pm}/Z^0$ go off-shell and 
the leptons they decay to are too soft to pass analysis cuts. 
Current searches are restricted to on-shell $W^{\pm}/Z^0$, so there is no bound 
from trilepton searches on degenerate Higgsinos. 
Exactly what $m_{\neut_2} - m_{\neut_1}$, $m_{\charpm_1} - m_{\neut_1}$ 
mass splitting the experiments are sensitive to is a very 
interesting question, but beyond the scope of this paper.

A more promising way to detect degenerate Higgsinos may be through 
monojet searches~\cite{monojet}. The initial quark/anti-quark in 
$q\bar q \ra \tilde{\chi}\tilde{\chi}$ production can emit hard radiation 
that recoils against the invisible portion of the event. 
Current monojet searches look for, among other physics, dark matter 
production after pairs of dark matter particles escape the detector. 
Re-interpreting the monojet bounds in terms of Higgsino pair production, 
we estimate that the current searches are sensitive to 
$\mu \sim 100\,\gev$~\cite{ushiggsinos}, competitive with the LEP
bound on charginos~\cite{LEPbound}. With more data, this bound may 
increase, slicing into the parameter space of Natural Supersymmetry. 
Additionally, there may be methods to optimize monojet searches 
for Higgsinos; the existing searches assume higher-dimensional 
contact operators whereas Higgsinos couple directly through 
electroweak gauge bosons.

If the Natural Supersymmetry expectations for $M_2$, $M_1$ are relaxed, 
it is also interesting to investigate how sensitive the LHC will be
to stop and sbottoms with light Higgsinos.  Two effects arise
from lowering $M_2$ and/or $M_1$:  the splittings between the
Higgsinos increase, and the gaugino content of the lightest
electroweakinos increases.  Since stop and sbottom decays to
Higgsino-like electroweakinos are dominated by the top Yukawa coupling,
we don't anticipate significant effects on the decay branching
fractions even when $M_{2,1}$ drop below $m_{\tilde{q}}$, 
opening up decays to gauginos.  The larger effect is the increase
in the splittings between the Higgsinos themselves.  
Clearly another interesting question is to probe how large the
splitting needs to be before the search strategies described here
become diluted by the additional energy from transitions between 
Higgsinos.  Mixing the light electroweakinos with some bino, wino,
or singlino is highly relevant for the possibility that the lightest
neutralino could be dark matter, but this is beyond the scope of this
paper.

Natural Supersymmetry may also lead to an unusual signal for 
the first and second generation squarks. One decay possibility 
for a first/second generation squark in Natural Supersymmetry is to 
a quark plus a Higgsino. As the first and second generation Yukawas 
are so small, the decay proceeds through the wino/bino fraction 
of the lightest neutralino and is therefore suppressed by $O(g\, v^2/M_{1,2})$. 
A second decay possibility is the three-body decay, 
$\tilde q \ra j + \tilde{t}_1 t$ or $j + \tilde{b}_1 b$ via an off-shell gluino. 
This option is suppressed by the gluino mass and three-body phase space, 
but comes with QCD coupling strength. Depending on the hierarchy of 
$M_3, M_2$ as well as the mass of the light squarks relative 
to the stops/sbottoms, the three-body decay fraction can be 
substantial.\footnote{The strength of the three-body mode also depends 
on the mass character of the gluino. For Dirac gluinos the suppression 
in the three-body mode is $m_{\tilde q}/M^2_3$ rather than $1/M_3$, 
making it much smaller.} First/second generations squark decays to 
$j\,\tilde{t}_1\,t$ or $j\,\tilde{b}_1 b$ would have several consequences 
that would be interesting to explore in more detail. Two obvious consequences 
are that the energy per final state particle would be lower because the 
squarks decay to multiple particles, and the decays would contain 
heavy-flavor jets not usually associated with 
first/second generation searches. 

Finally, while Natural Supersymmetry in a low energy effective theory 
is straightforward to define and quantify, issues of naturalness
become muddled as this is embedded into an ultraviolet completion.
The obvious issue is the that leading-log corrections to the
electroweak symmetry breaking scale can quickly become a poor
approximation if the renormalization group evolution is 
substantial.  For instance, ``radiative'' electroweak symmetry breaking
arises when $m_{H_u}^2$ is ``driven'' negative by its interaction with 
the stops, which clearly requires renormalization group improvement 
to determine the size of the contribution to electroweak symmetry breaking.
This is precisely why we considered $m_{\tilde{q}}$ and $\mu$ to be 
free parameters, since their separation may be much smaller than
the leading-log approximation suggests.  How this impacts the 
larger spectrum, particularly the gluino, becomes a highly
model-dependent question.  Nevertheless, we believe our analysis
has captured the essential physics of Natural Supersymmetry,
and we remain optimistic that it can be discovered with 
continued analyses at LHC\@.


\section*{Acknowledgments}

We thank Joe Lykken, Steve Martin, Steve Mrenna, and Gilad Perez for 
many valuable conversations. A.~Martin thanks Boston University for 
computing resources.
G.~Kribs thanks the Ambrose Monell Foundation for support while at the
Institute for Advanced Study.
G.~Kribs and A.~Menon are supported in part by the US Department of Energy under 
contract number DE-FG02-96ER40969. 


\appendix


\section{Realizing the observed Higgs mass and branching ratios of 
             Scenario I, II and III}
\label{sec:NMSSM}

In this Appendix, we consider the possibility of realizing Natural Supersymmetry
scenarios with $m_h \sim 125$~GeV\@.
In particular we consider the NMSSM model~\cite{Fayet:1974pd} where the 
Higgs sector of the MSSM is extended by including a gauge singlet.  
The superpotential has the form
\begin{equation}
W = W_{\rm Yuk} + \lambda \hat H_u \hat H_d \hat S + \frac{\kappa}{3} \hat S^3
\end{equation}
where $W_{\rm Yuk}$ are the usual Yukawa interactions and the hatted fields 
denote the chiral superfields. The corresponding soft supersymmetry breaking
terms are 
\begin{eqnarray}
V_{\rm soft} &= m_{H_u}^2 |H_u|^2 + m_{H_d}^2 |H_d|^2 + m_S^2 |S|^2 \nonumber \\
&~~~~~~~~~~~~~~~~~~~~~ + \lambda A_\lambda S H_u H_d + \frac{\kappa A_\kappa}{3} S^3.
\end{eqnarray}

In addition to the $D$-term contributions to the Higgs mass, the additional 
$\lambda\, v^2 \sin^2 2\beta$ contribution can help raise the Higgs mass 
above the $Z$-boson mass. On the other hand, solving the minimization conditions 
leads to the electroweak symmetry breaking condition of Eq.~(\ref{eq:ewsbtree}) 
with $\mu \to \lambda x$.\footnote{The additional minimization condition of the
singlet leads to a modified fine-tuning condition for the NMSSM\@. 
A detailed discussion of the fine-tuning in the generalized NMSSM-like models 
can be found in Refs.~\cite{Ellwanger:2011mu,Ross:2011xv,Ross:2012nr}.} 
To maximize the tree-level contributions to the Higgs mass, we need
both the NMSSM quartic contribution as well as the usual $D$-term contribution,
and thus small $\tan \beta$.  Small 
$\tan \beta$ typically enhances the hierarchy between $m_{H_u}, m_{H_d}$ and the
electroweak scale\footnote{For alternative NMSSM scenarios utilizing large $\tan\beta$, see Ref.~\cite{Badziak:2013bda}}.  Due to this tension between the Higgs mass and the 
hierarchy of scales, we consider $\tan \beta \in (1.5,2)$. However, we can 
still simultaneously realize the Natural Supersymmetry spectra in this paper
and the observed Standard Model Higgs mass in the NMSSM\@. 
Using NMSSMtools3.2.4~\cite{Ellwanger:2004xm}, we find $m_h \simeq 125$~GeV for 
the parameter space point $\tan \beta = 1.5$, $A_t = A_b = A_\tau = 0$, 
$m_{\tilde f}^2 = 700$~GeV, $M_1 = M_2 = M_3 = 2$~TeV, $\lambda = 0.7$, 
$\kappa = 0.67$, $A_\lambda = -60$~GeV, $A_\kappa = -200$~GeV and 
$\mu_{\rm eft} = 200$~GeV\@.  Also, for this parameter point the low energy 
precision and flavor observables are within 2$\sigma$ of their experimental
values.\footnote{As $\lambda$ and $\kappa$ are both somewhat large, 
these couplings may develop a landau pole before the GUT scale. 
The UV completion of such models can be realized in ``fat Higgs''-like
scenarios~\cite{Harnik:2003rs}, however a detailed study of this issue 
is beyond the scope of this paper.} For this point the neutralino masses are 
$m_{\neut} =
(197\mbox{GeV},227\mbox{GeV},416\mbox{GeV},1.98\mbox{TeV},1.99\mbox{TeV})$ 
and the chargino masses are 
$m_{\charpm} = (200\mbox{GeV},1.98\mbox{TeV})$. 
We checked that the branching ratios of the squarks into Higgsinos are 
within 1-2\% of the an MSSM model with similar sfermion and Higgsino masses.


\section{Search details}
\label{app:search}

For completeness, in the following we detail the important search criteria 
for each collaboration's particular search strategy that was 
used in this paper.  

\subsection*{CMS stops, semi-leptonic, 9.7 $\fbinv$ \cite{cms_stops}}

\noindent 
Object Id:
\begin{itemize}
\item jets, $p_T > 30\,\gev$, $|\eta_j| < 2.5$, anti-k$_T$, $R = 0.5$. Flavor tagging applied to all jets within $|\eta_j|  < 2.5$
\item electrons (muons), $p_T > 30\,\gev$, $|\eta_{\ell}| < 1.44 (2.1)$
\item leptons within $\Delta R = 0.4$ of a jet are removed
\end{itemize}
Basic cuts:
\begin{itemize}
\item $\slashchar{E}_T > 50\,\gev$
\item exactly 1 lepton passing the criteria above
\item 3 or more jets, with at least one $b$-tagged
\end{itemize}
Analysis:
\begin{itemize}
\item Events passing the basic selection cuts are binned according to the transverse mass of the MET + lepton system and the missing energy. Transverse mass is defined as 
\begin{equation}
m^2_{T, \slashchar{E}_T-\ell} = 2\,(\slashchar{E}_T p_{T, \ell}\ -\, \vec{\slashchar{p}}_T \cdot \vec{p}_{T, \ell})
\end{equation}
\item The bins are, in the format $(m_{T, min}, \slashchar{E}_{T, min})$: \\
 $(150\,\gev, 100\,\gev)$,$(120\,\gev, 150\,\gev)$, $(120\,\gev, 200\,\gev)$, $(120\,\gev, 250\,\gev)$, \\ $(120\,\gev, 300\,\gev)$, $(120\,\gev, 350\,\gev)$,$(120\,\gev, 400\,\gev)$    
\item The bins are not exclusive, so the bin with the best limit at a given ($m_{\tilde t_1}, m_{\neut_1}$) point is used.
\end{itemize}

\subsection*{ATLAS stops, semi-leptonic, 20.7 $\fbinv$ \cite{atlas_stops_semi}}

\noindent
Object Id:
\begin{itemize}
\item jets, $p_T > 20\,\gev$, $|\eta_j| < 2.5$, anti-k$_T$, $R = 0.4$. Flavor tagging applied to all jets within $|\eta_j|  < 2.5$
\item electrons (muons), $p_T > 10\,\gev$, $|\eta_{\ell}| < 2.7 (2.4)$
\item any jet within $\Delta R = 0.2$ of an electron is removed
\item subsequently, any leptons within $\Delta R = 0.4$ of a jet are removed
\end{itemize}
Basic cuts:
\begin{itemize}
\item exactly 1 lepton, which must have $p_{T, \ell} > 25\,\gev$
\item 4 or more jets, at least one of which is $b$-tagged. The four hardest jets must satisfy $p_T > 80\,\gev, 60\,\gev, 40\,\gev, 25\,\gev$ respectively.
\end{itemize}
Analysis:
\begin{itemize}
\item Most channels require top reconstruction, done as follows: the closest pair of jets (in $\Delta R$) that satisfy $m_{jj} > 60\,\gev$ are dubbed a 'W-candidate'. This candidate is combined with the nearest jet (again in $\Delta R$). For the resulting three-jet system to be considered a successful top-candidate, $130\,\gev < m_{jjj} < 205\,\gev$ is required.
\item Other analysis cuts include: the transverse mass of the $\slashchar{E}_T - \ell$ system, the missing energy, the $\slashchar{E}_T$-significance -- defined as $\slashchar{E}_T/\sqrt{H_{T,j_{1-4}}}$, and the $\Delta \phi$ between the missing energy (transverse) vector and the leading two jets. 
\item In the channels designed to be sensitive to the highest stop masses, two other variables are included $a m_{T,2}$ and $m^{\tau}_{T,2}$, both of which are slight variants on the $m_{T,2}$ variable~\cite{Lester:1999tx}. In $m_{T,2}$, as in these variations, the visible part of the event is divided into two, and all partitions of the missing energy are scanned over. The difference between $m_{T,2}, a m_{T,2}$, and $m^{\tau}_{T,2}$ lie in whether all the visible particles are used, or only some of them. In $a m_{T,2}$, only the leading light jet, lepton, and highest weight $b$-jet are taken as the visible part of the event, while in $m^{\tau}_{T2}$ only the leading lepton and leading light jet are used.
\item The channels dedicated to $\tilde t \rightarrow t + \neut$ are:
\begin{enumerate}
\item 1 top candidate, alone with $\Delta \phi_{\slashchar{E}_T-j_1} > 0.8$, $\Delta \phi_{\slashchar{E}_T-j_2} > 0.8$, $\slashchar{E}_T > 100\,\gev$, $\slashchar{E}_T$ signif. $> 13,\,  M_{T} > 60\,\gev$. Events passing this selection are then separated into 12 finer bins according to their $M_{T}$ and $\slashchar{E}_T$: \\
$M_{T} \in \{60 -90\,\gev, 90-120\,\gev$,\\$~~~~~~~120 -140\,\gev, >140\,\gev\}$\\
$\slashchar{E}_T \in \{100-125\,\gev, 125-150,\,\gev$\\$~~~~~~~> 150\,\gev\}$. 
\item 1 top candidate, along with $\Delta \phi_{\slashchar{E}_T-j_2} > 0.8$, $\slashchar{E}_T > 200\,\gev$, $\slashchar{E}_T$ signif. $> 13,\,  M_{T} > 140\,\gev$, $a m_{T,2} > 170\,\gev$
\item 1 top candidate, along with $\Delta \phi_{\slashchar{E}_T-j_1} > 0.8$, $\Delta \phi_{\slashchar{E}_T-j_2} > 0.8$, $\slashchar{E}_T > 275\,\gev$, $\slashchar{E}_T$ signif. $> 11,\, M_{T} > 200\,\gev,\ a m_{T,2} > 175\,\gev, m^{\tau}_{T,2} > 80\,\gev$
\end{enumerate}
\item The analysis contains three channels aimed at the $\tilde t \rightarrow b+\charpm$ final state. In these channels no top candidate is required. Instead there are stronger requirements on the $p_T$  and multiplicity of the $b$-jets, and an additional cut on $m_{eff}$, defined as the scalar sum of the $p_T$ of all jets with $p_T > 30\,\gev$ summed together with the $\slashchar{E}_T$ magnitude and $p_{T, \ell}$
\end{itemize}

\subsection*{ATLAS stops, fully hadronic, 20.5 $\fbinv$ \cite{atlas_stops_had}}

\noindent
Object Id:
\begin{itemize}
\item jets, $p_T > 20\,\gev$, $|\eta_j| < 4.5$, anti-k$_T$, $R = 0.4$. Flavor tagging applied to all jets within $|\eta_j|  < 2.5$
\item electrons (muons), $p_T > 10\,\gev$, $|\eta_{\ell}| < 2.7 (2.4)$
\item any jet within $\Delta R = 0.2$ of an electron is removed
\item subsequently, any leptons within $\Delta R = 0.4$ of a jet are removed
\end{itemize}
Basic cuts:
\begin{itemize}
\item zero leptons passing the above criteria
\item $\slashchar{E}_T > 130\,\gev$
\item 6 or more jets, where jets satisfy $p_T > 33\,\gev$, $|\eta_j| < 2.8$. The leading two jets must have $p_T > 80\,\gev$, and at least two jets are $b$-tagged.
\end{itemize}
Analysis:
\begin{itemize}
\item 2 three-jet clusters are formed from from the list of jets as follows: the three jets that are closest in the $\phi-\eta$ plane are taken as one such cluster, removed from the list, then the process is repeated to extract the second group. The mass of these three-jet clusters is required to lie within $80\,\gev < m_{jjj} < 270\,\gev$ in order to select events with two hadronic tops in the final state.
\item The transverse mass of the $\slashchar{E}_T-b$ system, where the $b$ closest in $\Delta \phi$ is used is required to be $> 175\,\gev$ to remove leptonic $\bar t t$ background
\item  $\Delta \phi_{\slashchar{E}_T-j} > 0.2\pi$, where $\Delta \phi_{\slashchar{E}_T j}$ is the angle between the missing energy vector and the closest jet. This cut is designed to remove backgrounds from mismeasured jets.
\item The remaining events are binned according to $\slashchar{E}_T$: $\slashchar{E}_T > 200\,\gev, >\, 300\,\gev$ and $\slashchar{E}_T > 250\,\gev$. Only the strongest limit at a given ($m_{\tilde t_1}, m_{\neut_1}$) point is used.
\end{itemize}

\subsection*{CMS sbottoms, multi-$b$ + $\slashchar{E}_T$, 11.7\,$\fbinv$ \cite{Chatrchyan:2013lya}}

\noindent 
Object Id:
\begin{itemize}
\item jets, $p_T > 50\,\gev$\footnote{This requirement is scaled down to $37\,\gev$, or $43\,\gev$ for events with low-$H_T$.}, $|\eta_j| < 3.0$, anti-k$_T$, $R = 0.5$. Flavor tagging applied to all jets within $|\eta_j|  < 2.5$
\item electrons (muons), $p_T > 10\,\gev$, $|\eta_{\ell}| < 2.4 (2.1)$
\item any jet within $\Delta R = 0.4$ of a lepton is removed
\end{itemize}
Basic cuts:
\begin{itemize}
\item zero leptons
\item at least 2 jets. The hardest  jet must lie within $|\eta| < 2.5$ and the leading two jets must have $p_T > $ twice the nominal jet $p_T$ requirement. Nominally this is $> 100\,\gev$ but for events with low-$H_T$ this cut may be softer. Events with high-$p_T$ jets (i.e. passing nominal jet criteria) at $|\eta| > 3.0$ are vetoed.
\item $H_T > 275\,\gev$, where $H_T$ is the scalar sum of the $p_T$ of all jets in the events.
\end{itemize}
Analysis cuts:
\begin{itemize}
\item All visible objects in the event are grouped into two mega-jets, following the criteria given in Ref.~\cite{}. The degree to which the two megajets balance each other is captured by the variable $\alpha_T = \frac{E_{T,2}}{M_{T,jj}}$, the fraction of the transverse energy of the subleading (in $p_T$) megajet relative to the transverse mass of the megajet pair. Requiring $\alpha_T > 0.55$ greatly suppresses multijet QCD backgrounds.
\item Events surviving the $\alpha_T$ cut are categorized according to the jet and $b$-jet multiplicities, then binned in $H_T$.
\item The jet multiplicity categories are $N_{jet} = 2-3$, and $N_{jet} = 4^+$. Within each jet multiplicity category, $N_b = 0,1,2,3, (4)$ is considered (obviously  $N_b = 4$ is only considered in the $N_{jet} >= 4$ class). For a given $(N_{jet}, N_b)$, the $H_T$ is binned as $[275-325\,\gev], [325-375\,\gev], [375-475\,\gev],$, $[475-575\,\gev], [575-675\,\gev], [675-775\,\gev]$, $[775-875\,\gev]$ and $> 875\,\gev$\footnote{For the samples with the highest $b$-multiplicity, only three $h_T$ bins are used, $[275-325\,\gev], [325-375\,\gev], >375\,\gev$.}.
\item For the direct sbottom search, only the $N_{jet} = 2-3, N_b = 0, 1$ categories are used to set limits. Since the $H_T$ bins are orthogonal, all $H_T$ bins across both categories are taken together to form a combined limit.
\end{itemize}

\subsection*{ATLAS sbottoms, multi-$b$ + $\slashchar{E}_T$, 12.8$\fbinv$ \cite{atlas_sbottoms}}

\noindent 
Object Id:
\begin{itemize}
\item jets, $p_T > 20\,\gev$, $|\eta_j| < 2.8$, anti-k$_T$, $R = 0.4$. Flavor tagging applied to all jets within $|\eta_j|  < 2.5$
\item electrons (muons), $p_T > 10\,\gev$, $|\eta_{\ell}| < 2.7 (2.4)$
\item any jet within $\Delta R = 0.2$ of an electron is removed
\item subsequently, any leptons within $\Delta R = 0.4$ of a jet are removed
\end{itemize}
Basic cuts:
\begin{itemize}
\item zero leptons
\item 2 or more jets, with 2 or more $b$-tags
\item $\slashchar{E}_T > 150\,\gev$
\end{itemize}
Analysis:
\begin{itemize}
\item After basic selection, 3 event categories are set up, each with slightly different requirements. The categories are not exclusive: \\
\begin{enumerate}
\item leading jet $p_T > 150\,\gev$, subleading jet $p_T > 50\,\gev$, no other jets with $p_T > 50\,\gev$. Both the leading two jets must be tagged as $b$ jets. To reduce multijet QCD, $\Delta \phi_{\slashchar{E}_T- j_2} > 0.4$ and $\slashchar{E}_T/m_{eff} > 0.25$ are required. Here $m_{eff}$ is the scalar sum of the missing energy and the $p_T$ of the hardest three jets satisfying basic jet requirements (meaning they must be harder than $20\,\gev$ only) ,$m_{eff}  = \slashchar{E}_T + \sum_{i=1}^{3} p_{T,j_i}$. Within this category, events are further binned according to their contra-transverse mass, see Ref.~\cite{Tovey:2008ui} for definition.
\item similar to category 1.) but the $p_T$ requirements are adjusted to $>200\,\gev, >60\,\gev$  for the leading and subleading jets. The leading two jets still must be flavor tagged, and the $\Delta \phi_{\slashchar{E}_T- j_2}$ and the $\slashchar{E}_T/m_{eff} $ are unchanged.
\item More than 2 jets are required with the leading jet having $p_T > 130\, \gev$. The two subleading jet must have $p_T > 30\,\gev$, but below $110\,\gev$. Unlike the previous categories, the first two categories, the leading jet is not required to be a $b$-jet. Instead the leading jet must have light flavor (it is anti-tagged), while the subleading two jets must be tagged as $b$-jets. The $\Delta \phi_{\slashchar{E}_T- j_2}$ and the $\slashchar{E}_T/m_{eff} $ are unchanged, but there is an additional requirement that the scalar sum of the $p_T$ of all jets beyond the leading three is small, $< 50\,\gev$. This category is divided into two subcategories with different $\slashchar{E}_T $ and $p_{T,j_1}$ requirements.
\end{enumerate}
\end{itemize}


\end{document}